\documentclass[aps,prx,twocolumn,english,balance,superscriptaddress,floats,showpacs,letter, prb]{revtex4-2}
\usepackage[latin9]{inputenc}
\usepackage{amsmath}
\usepackage{amssymb}
\usepackage{graphicx}
\usepackage{esint}
\usepackage{subfigure}
\usepackage{multirow}
\usepackage{mathtools}
\usepackage{xcolor}
\usepackage{gensymb}

\makeatletter

\newcommand{\beq}{\begin{equation}}
\newcommand{\eeq}{\end{equation}}
\newcommand{\bea}{\begin{eqnarray}}
\newcommand{\eea}{\end{eqnarray}}
\newcommand{\bwt}{\begin{widetext}}
\newcommand{\ewt}{\end{widetext}}
\@ifundefined{textcolor}{}
{%
 \definecolor{BLACK}{gray}{0}
 \definecolor{WHITE}{gray}{1}
 \definecolor{RED}{rgb}{1,0,0}
 \definecolor{GREEN}{rgb}{0,1,0}
 \definecolor{BLUE}{rgb}{0,0,1}
 \definecolor{CYAN}{cmyk}{1,0,0,0}
 \definecolor{MAGENTA}{cmyk}{0,1,0,0}
 \definecolor{YELLOW}{cmyk}{0,0,1,0}
}

\newcommand{\eps}{\epsilon}
\newcommand{\bk}{\mathbf{k}}
\newcommand{\bq}{\mathbf{q}}
\newcommand{\bp}{\mathbf{p}}
\newcommand{\bg}{\mathbf{g}}
\newcommand{\br}{\mathbf{r}}
\newcommand{\bL}{\mathbf{L}}
\newcommand{\bK}{\mathbf{K}}
\newcommand{\bR}{\mathbf{R}}

\newcommand{\fvec}[1]{\boldsymbol{#1}}

\makeatother

\begin{document}

\title{Towards the hidden symmetry in Coulomb interacting twisted bilayer graphene: renormalization group approach}

\author{Oskar Vafek}
\email{vafek@magnet.fsu.edu}
\affiliation{National High Magnetic Field Laboratory, Tallahassee, Florida, 32310, USA}
\affiliation{Department of Physics, Florida State University, Tallahassee, Florida 32306, USA}

\author{Jian Kang}
\email{jkang@suda.edu.cn}
\affiliation{School of Physical Science and Technology {\normalfont \&} Institute for Advanced Study, Soochow University, Suzhou, 215006, China}

\begin{abstract}
We develop a two stage renormalization group which connects the continuum Hamiltonian for twisted bilayer graphene at length scales shorter than the moire superlattice period to the Hamiltonian for the active narrow bands only which is valid at distances much longer than the moire period. In the first stage, the Coulomb interaction renormalizes the Fermi velocity and the interlayer tunnelings in such a way as to suppress the ratio of the same sublattice to opposite sublatice tunneling, hence approaching the so-called chiral limit. In the second stage, the interlayer tunneling is treated non-perturbatively. Via a progressive numerical elimination of remote bands the relative strength of the one-particle-like dispersion and the interactions within the active narrow band Hamiltonian is determined, thus quantifying the residual correlations and justifying the strong coupling approach in the final step. We also calculate exactly the exciton energy spectrum from the Coulomb interactions projected onto the renormalized narrow bands. The resulting softening of the collective modes marks the propinquity of the enlarged (``hidden'') $U(4)\times U(4)$ symmetry in the magic angle twisted bilayer graphene.
\end{abstract}

\maketitle

It has been known for some time that the electron-electron Coulomb interactions cause an upward renormalization of the Fermi velocity, $v_F$, upon approaching the charge neutrality point (CNP) of mono-layer graphene\cite{GeimNovoselovNatPhys2011,GeimNovoselovPNAS2013,Gonzalez1994,VafekPRL2007,SheehyPRL2007,BorghiSSC2009,BarnesPRB2014}. Such momentum dependent steepening of the Dirac cone depends on the graphene's dielectric environment and is weaker for stronger dielectrics, but even for hexagonal boron nitride (hBN) encapsulated devices the increase can be\cite{GeimNovoselovPNAS2013} $\sim 10-15\%$.
Such a small change in $v_F$ would be of limited interest if it weren't for the recent explosion of research into the magic angle~\cite{BMModel} twisted bilayer graphene (TBG)~\cite{Pablo1,Pablo2,Cory1,David,Young,Cory2,Dmitry1,Yazdani,Ashoori,Dmitry2,Eva,Yazdani2,Shahal,Young2,Stevan,YuanCao2020,Young3,Xu,KangVafekPRX,LiangPRX1,Senthil1,Leon1,FanYang,Kuroki,Kivelson,LiangPRX2,Louk,Guo,GuineaPNAS,BJYangPRX,Bernevig1,Leon2,Dai1,FengchengSC,Fernandes1,Qianghua,Grisha,Stauber,KangVafekPRL,Bruno,Senthil2,SenthilC3,Ashvin1,Cantele,Cenke,MacDonald,Thomson,Zaletel1,Guinea2,Senthil3Ferro,Ashvin2,Sau,Zaletel2,Zaletel3,Chubukov,Dai2,YiZhang,KangVafekPRB,Ziyang,Chubukov2,Roy,Fernandes2,Rahaul,Fengcheng,Kaxiras2019,SenthilTop,LeonReview,Lucile,Zaletel4}, where the experiments show extremely strong sensitivity of the correlated electron phenomena to the twist angle $\theta$. Even a $\sim 5\%$ change of $\theta$ away from the optimal (magic) value has been reported to produce at least a factor of $2$ reduction\cite{Young2,YuanCao2020} of the superconducting $T_c$, with even stronger suppression of the correlated insulator states\cite{Young2}.

The strong band structure sensitivity is due to the dependence on the dimensionless parameters $w_{0,1}/v_Fk_\theta$, where $w_{0}$ and $w_1$ parameterize the interlayer tunneling energy in the $AA$ and $AB$ regions respectively, and where the momentum displacement of the Dirac cones is given by $k_\theta=2k_D\sin\frac{\theta}{2}$, $k_D=4\pi/3a_0$, $a_0\approx 0.246$nm (in $\hbar=1$ units)~\cite{BMModel}. Therefore, at a fixed magic $\theta$, even a $\sim 10\%$ percent difference in $v_F$ alone would be sufficient to de-tune the system from the optimal flat band condition. As such, if neither of $w_j$ renormalized due to Coulomb interactions, but only $v_F$ did, the magic angle condition would depend on whether the TBG was encapsulated in the hBN,
or only from one side,
because the different dielectric environments would produce a different strength of Coulomb interactions, former with a dielectric constant\cite{Hunt2017,KangVafekPRL} $\eps_{hBN}\approx 4.4$ and the latter with $\eps\approx(1+\eps_{hBN})/2= 2.7$. The difference in the $v_F$, and therefore the magic angle, would then be within the sensitivity of the correlated insulating states; no such dependence of the magic angle on the partial or complete encapsulation has been reported.

\begin{figure}[t]
	\begin{minipage}[b]{0.4\linewidth}
		\centering
		\subfigure[\label{Fig:SchematicLattice}]{\includegraphics[width=\textwidth]{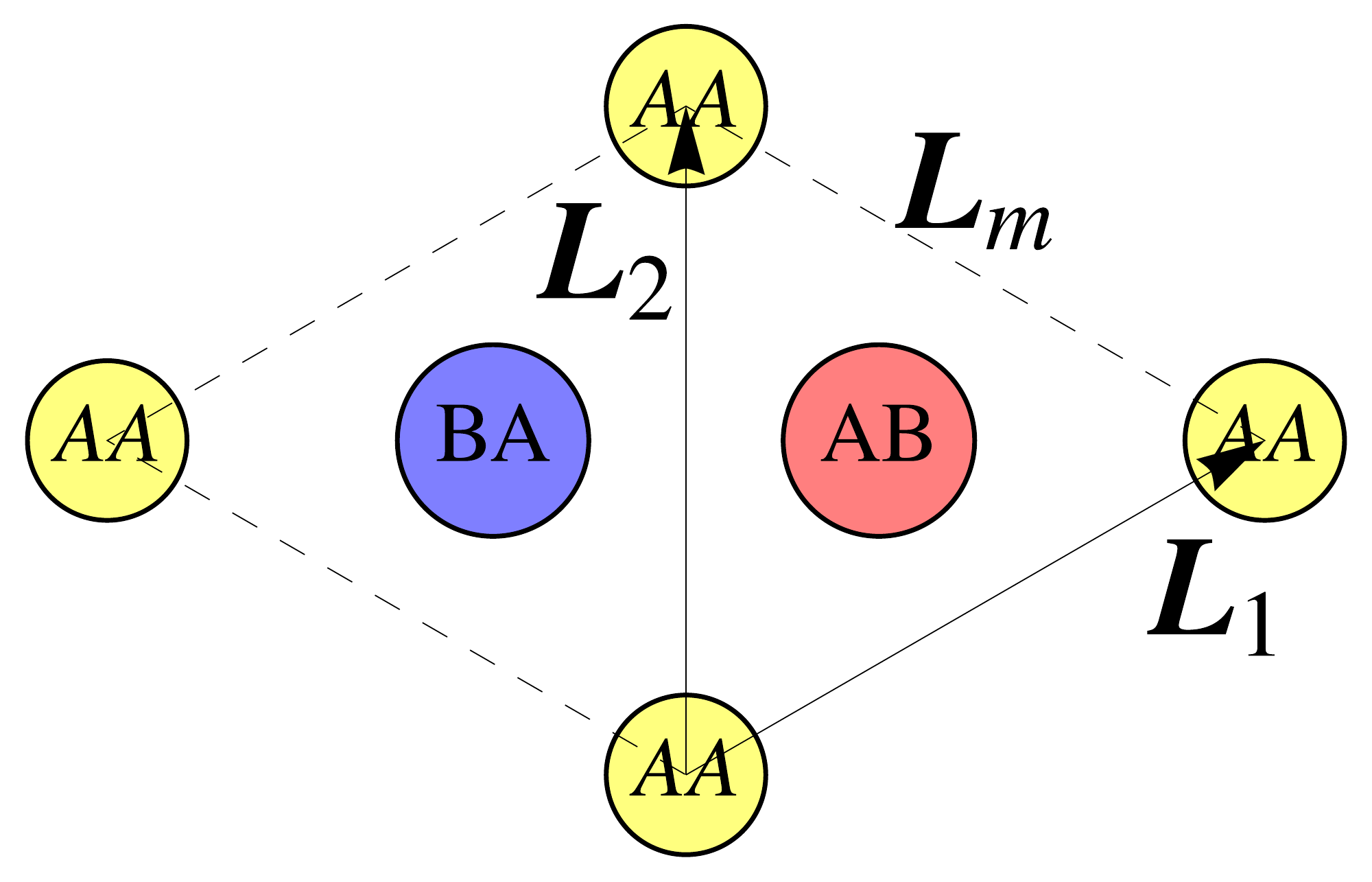}}\vspace{0.5cm}
		\subfigure[\label{Fig:SchematicMomP}]{\includegraphics[width=\textwidth]{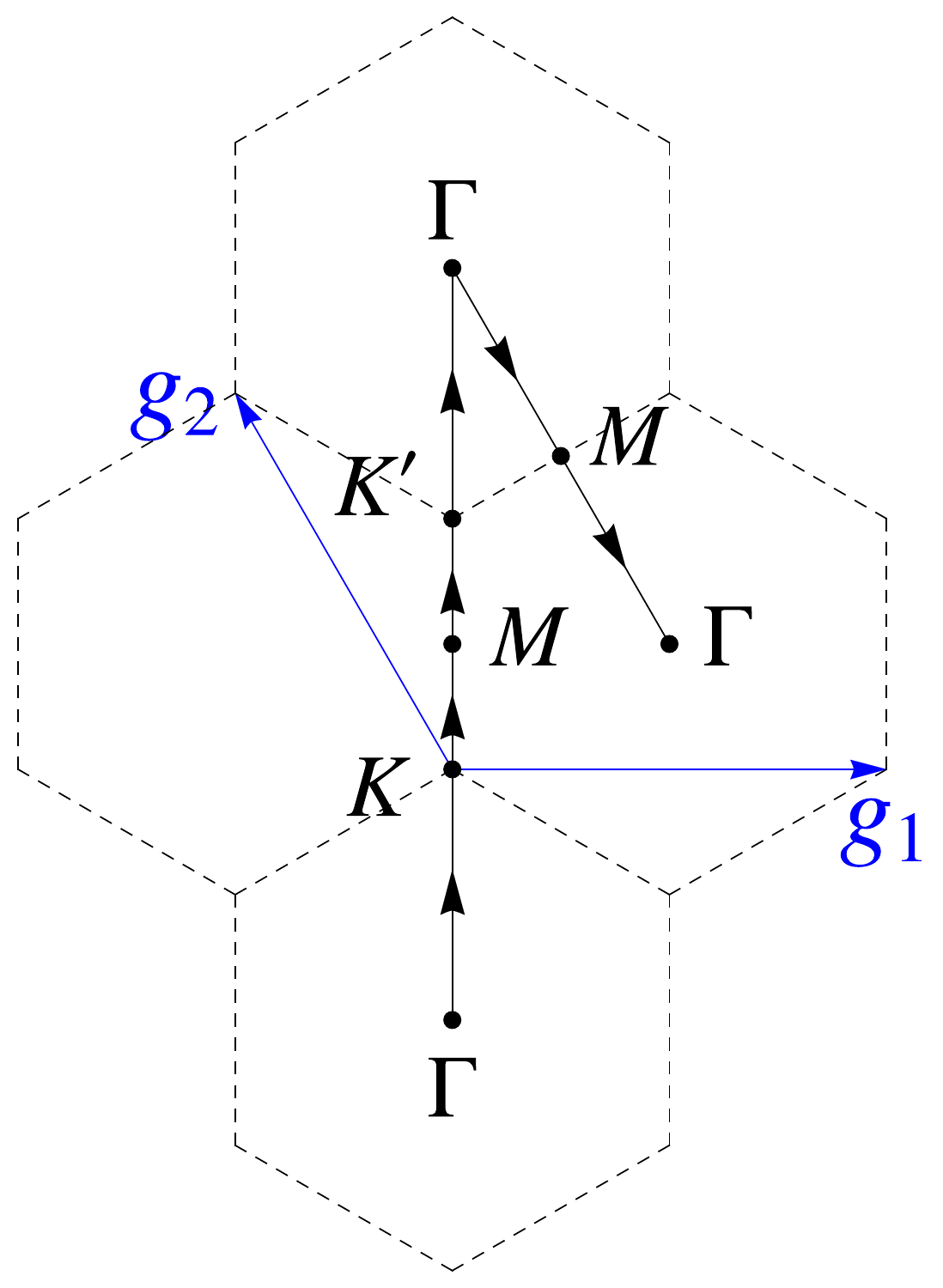}}
	\end{minipage}
	\hspace{0.2cm}
	\begin{minipage}[b]{0.55\linewidth}
		\centering
		\subfigure[\label{Fig:SchematicRG}]{\includegraphics[width=\textwidth]{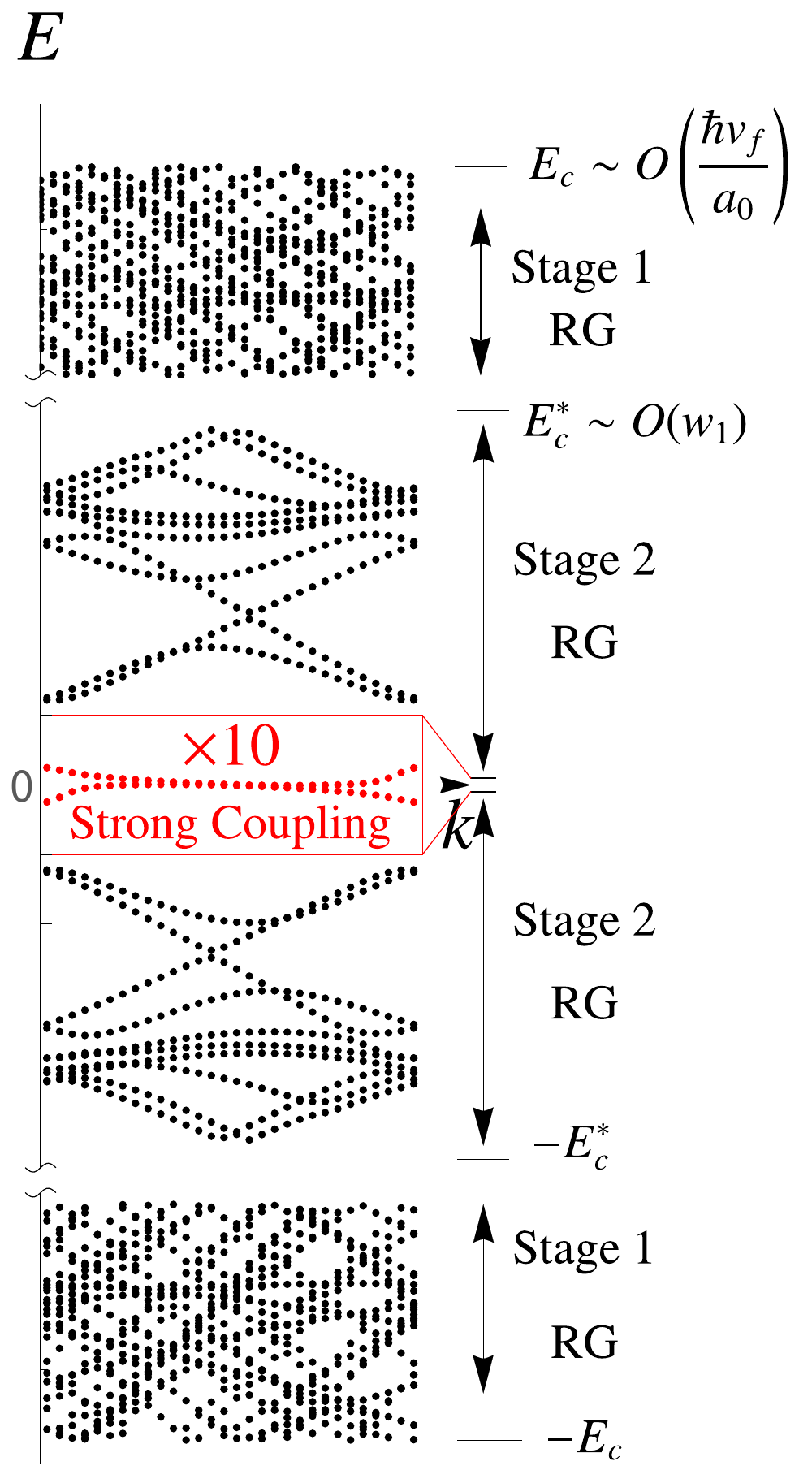}}
	\end{minipage}
    \caption{(a) Moire lattice with lattice spacing $L_m$. (b) Moire Brillouin zone. (c) Schematic illustration of the two stage RG procedure for arriving at the strong coupling limit. In the stage 1, both the Coulomb interaction and the moire  potential are perturbative, in the stage 2 only the Couloumb interaction is. In the final step, when only the narrow bands (red) remain, the interaction is the largest scale.}
	\label{Fig:schematics}
\end{figure}

\begin{figure}[tbp]
	\centering
	\subfigure[\label{Fig:nonchiralunscaled}]{\includegraphics[width=0.49\columnwidth]{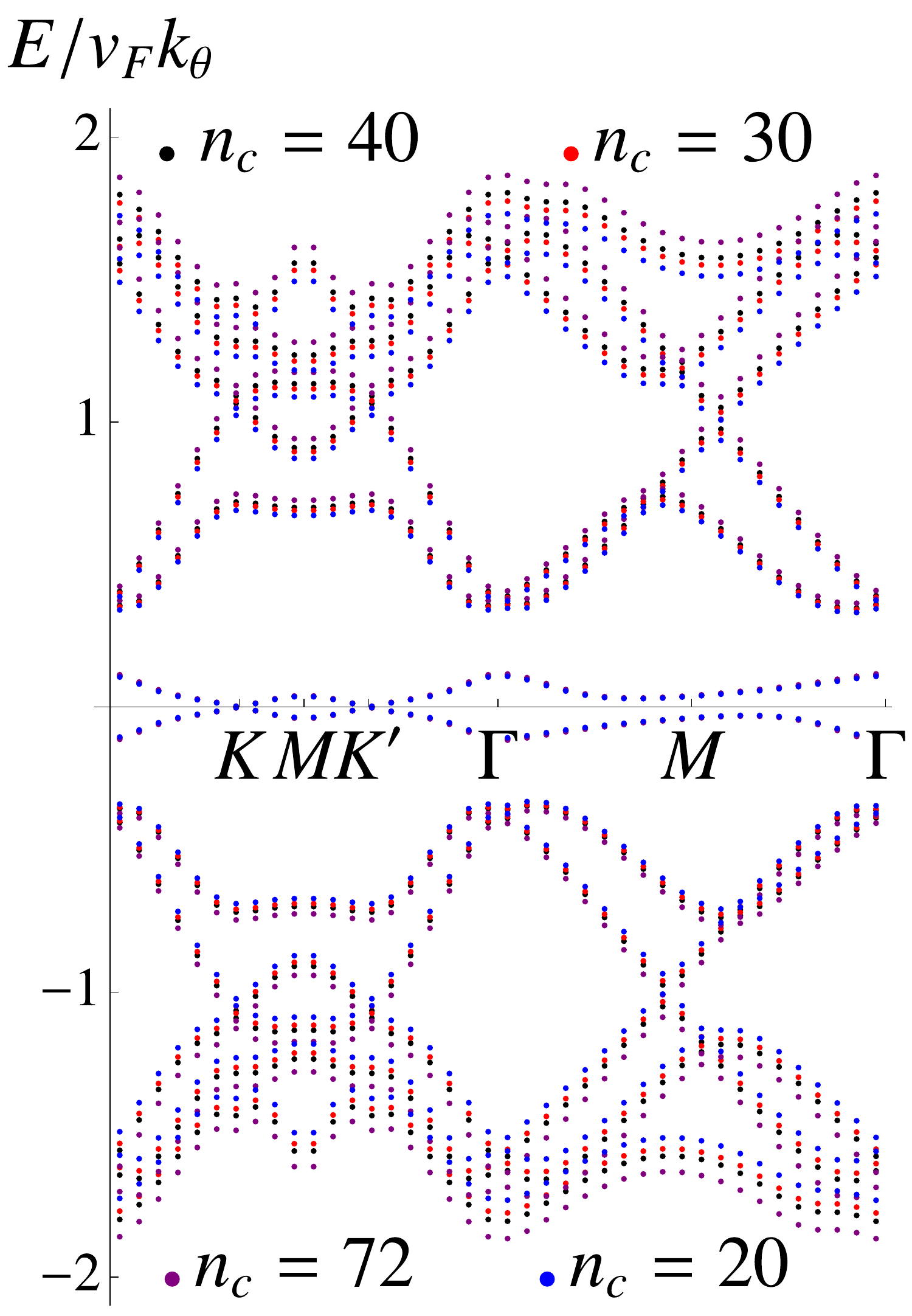}}
	\subfigure[\label{Fig:nonchiralunscaled}]{\includegraphics[width=0.49\columnwidth]{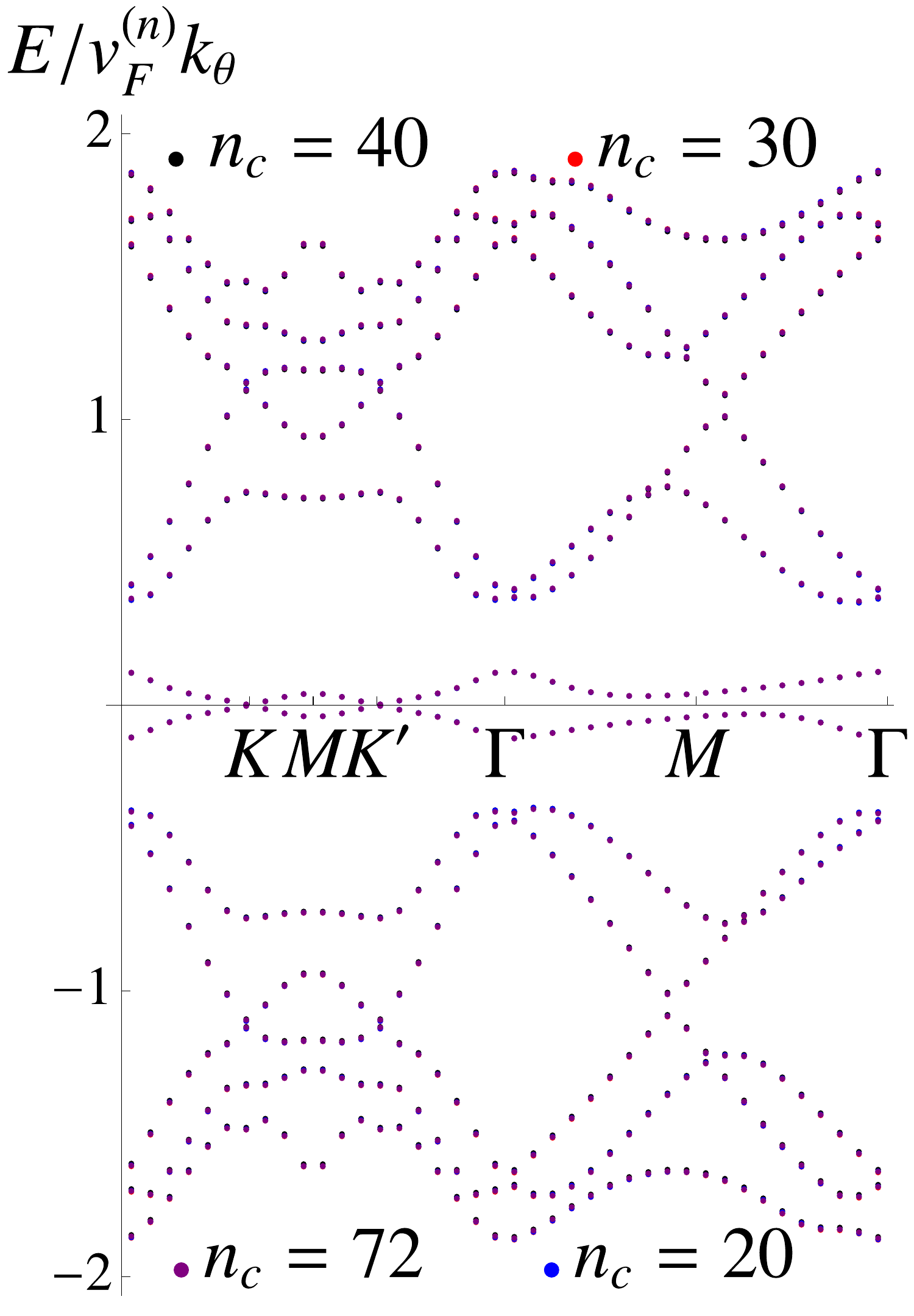}}
	\caption{(a) Low energy spectra after $n_c-5$ steps of the stage 2 RG for $n_c=72$(purple), $40$ (black), $30$ (red), $20$ (blue). At $n_c$, each starts with the same Fermi velocity, $v_F$, in the BM model at $w_1/v_Fk_\theta=0.5$, but with $w_0/w_1=0.83$ (purple), $0.805$ (black), $0.787$ (red) and $0.768$ (blue). The values are chosen based on the dielectric constant $\eps=4.4$ and scaling in Eq.(\ref{Eqn:chiral limit}) and the cutoff energies set by the $n$th band maxima at $n_c=72$. (b) Results of the panel (a) rescaled by $v^{(n)}_F=v_F/(1+\frac{e^2}{4\eps v_F}\ln\frac{E_c}{E_c^*})$ for $E_c$ set by the band maximum at $n_c=72$, demonstrating the scaling collapse and thus independence of the results of stage 2 RG on $E^*_c$.}
	\label{Fig:nonchiralscaling}
\end{figure}

Here we develop a renormalization group (RG) approach to the Coulomb interactions in the twisted bilayer graphene and show that $w_1$ renormalizes in precisely such a way as to compensate for the growth of $v_F$ making the magic angle largely insensitive to the effective dielectric constant $\eps$. Interestingly, we find that $w_0$ does not renormalize due to Coulomb interactions. Therefore, the ratio $w_0/w_1$ shrinks and the system flows closer to the chiral limit described by Tarnopolsky, Kruchkov and Vishwanath\cite{Grisha}.
As illustrated in the Fig.~\ref{Fig:schematics}c, the flow from a high energy (with the UV cutoff $E_c$), where the Coulomb interaction and $w_{0,1}$ are perturbative, to a low energy of the narrow bands where neither is, crosses over to a regime where the effects of $w_{0,1}$ become non-perturbative, but the Coulomb interaction is still perturbative. This happens at the energy scale $E^*_c\sim \mathcal{O}(w_1)$, marking the beginning of the second stage of our RG; the band structure scaling collapse in Fig.\ref{Fig:nonchiralscaling} shows that the $2^{nd}$ stage seamlessly connects to the $1^{st}$ stage even if $E^*_c$ changes. In the $2^{nd}$ stage, we numerically integrate out the two most remote bands, one above and one below the CNP, rotate the remaining states to diagonalize the renormalized kinetic energy and re-express the interaction in terms of the rotated states, iterating the procedure until we reach the narrow bands. If the resulting narrow bands bandwidth (or, more precisely the root-mean-square of the renormalized kinetic energy dispersion) is much smaller than the interaction (or more precisely, the particle-hole charge gap), as we find it is near the magic angle, the final step is treated non-perturbatively in the Coulomb interaction i.e. by solving the interaction-only problem (strong coupling limit) and then treating the renormalized kinetic energy terms as a perturbation.

\begin{figure}[t]
	\centering
	\includegraphics[width=\columnwidth]{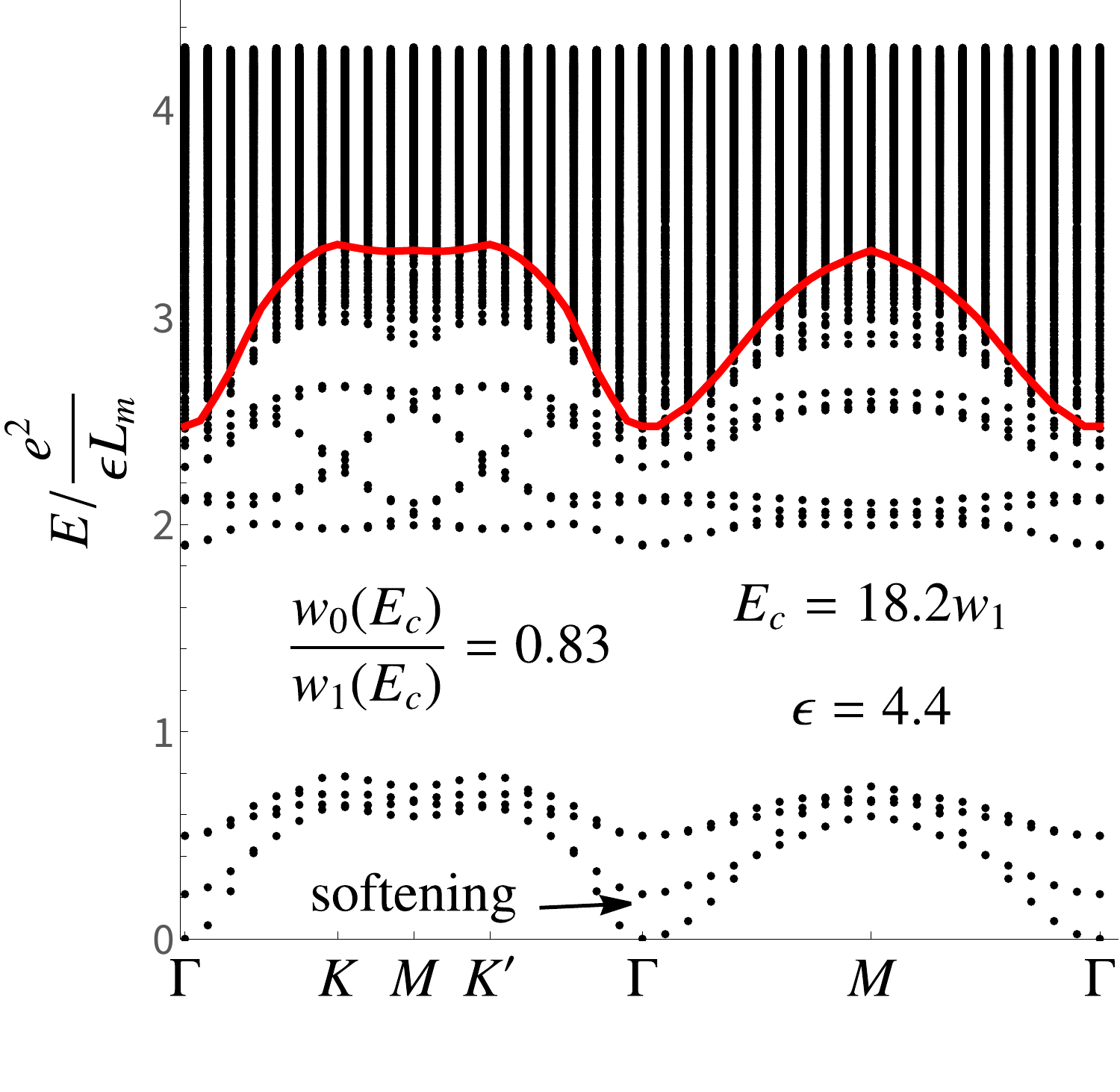}
	\caption{The strong coupling exciton spectrum after stage 1 and 2 RG, starting the stage 1 with $E_c = 18.2 w_1$ corresponding to $2$eV for $w_1=110$meV, $w_1/(v_F k_{\theta}) = 0.586$ (magic angle), and the initial $w_0/w_1 = 0.83$. The branch that becomes gapless at $\fvec \Gamma$ corresponds to 4 Goldstone modes of $U(4)$ spin/valley ferromagnet with quadratic dispersion. Another branch, emphasized by the arrow, softens during the RG, eventually also becoming gapless in the chiral limit, with the total of 8 Goldstone modes of $U(4)\times U(4)$ ferromagnet. The red curve is the onset of the particle-hole continuum.}
	\label{Fig:Exciton}
\end{figure}

The condition $w_0=0$, and thus the chiral limit\cite{Grisha,Niu2020,Becker2020}, was previously thought to be unrealistic and the value $w_0/w_1\sim 0.8$ was taken from DFT-like calculations \cite{NamKoshino2017,LiangPRX1,Kaxiras2019}. Our results (\ref{Eqn:RG magic angle}-\ref{Eqn:chiral limit}) show that for Coulomb interacting system, the chiral limit becomes exact near the CNP in the limit $E_c/w_1\rightarrow \infty$, albeit approaching logarithmicaly.
This has important consequences for the effective residual interaction in the narrow band, because of the increased sublattice polarization of the narrow band wavefuctions\cite{Zaletel3}. We find additional enhancement of the sublattice polarization after the $2^{nd}$ stage, as well as steepening of the Wilson loop eigenvalues\cite{Bernevig1}, indicating additional approach to the chiral limit during the $2^{nd}$ stage RG.
The dominant part of the Coulomb interaction Hamiltonian projected onto perfectly sublattice polarized chiral limit narrow bands is invariant under a larger symmetry, $U(4)\times U(4)$, than for $w_0/w_1\neq 0$, $U(4)$, when particle-hole (p-h) symmetry\cite{Bernevig1} is exact\cite{Zaletel3}. This symmetry enhancement enlarges the manifold of nearly degenerate correlated states\cite{Zaletel3}.
Our exact calculation of the collective mode spectrum in the strong coupling limit indeed shows not only 4 Goldstone bosons associated with the $U(4)$ spin-valley ferromagnetism\cite{KangVafekPRL,Zaletel3}, but also a softening of 4 additional collective modes, indicating the approach to the $U(4)\times U(4)$ ferromagnet\cite{Zaletel3} with its $8$ Goldstone bosons (see Fig.~\ref{Fig:Exciton}).

We begin with the Hamiltonian $H = H_{kin} + V_{int}$
where
\begin{eqnarray}\label{Eqn:starting Hamiltonian kin}
H_{kin} &=& \int d^2\br \chi_\sigma^\dagger(\br)
\left(\begin{array}{cc}\hat{H}_{BM}& 0 \\ 0 & \hat{H}^*_{BM}\end{array}\right)\chi_\sigma(\br)\\
V_{int} &=& \frac{1}{2}\int d^2\br d^2\br' V(\br-\br') \chi_\sigma^\dagger(\br) \chi_{\sigma'}^\dagger(\br')\chi_{\sigma'}(\br')
\chi_{\sigma}(\br)
\label{Eqn:starting Hamiltonian}
\end{eqnarray}
where $\chi_\sigma^\dagger=(\psi^\dagger_\sigma,\phi^\dagger_\sigma)$ creates an electron in valley $\bK$ ($\bK'$) for its upper (lower) component, and the repeated spin-$\frac{1}{2}$ indices $\sigma$ are summed.
The Bistritzer-MacDonald\cite{BMModel} (BM) continuum Hamiltonian \cite{LiangPRX1,Senthil1,Bernevig1,Grisha,Leon1}  for twist angle $\theta$ is
\begin{eqnarray}
\hat{H}_{BM} &=& \left(\begin{array}{cc} v_F\sigma_{\frac{\theta}{2}}\cdot \bp & T(\br) \\
T^\dagger(\br) & v_F\sigma_{-\frac{\theta}{2}}\cdot \bp
\end{array}\right),
\label{Eqn:BM}
\end{eqnarray}
where the twisted Pauli matrices acting on the sublattice indices are $\sigma_\frac{\theta}{2}=e^{-\frac{i}{4}\theta\sigma_z}(\sigma_x,\sigma_y)e^{\frac{i}{4}\theta\sigma_z}$, $\bq_1=k_\theta(0,-1)$, $\bq_{2,3}=k_\theta\left(\pm\frac{\sqrt{3}}{2},\frac{1}{2}\right)$. The interlayer hopping $T(\br)=\sum_{j=1}^3 T_j e^{-i\bq_j\cdot\br}$ is controlled by two parameters $w_{0,1}$ via
\begin{eqnarray}
T_{j+1}=w_0 1_2+w_1\left(\cos\left(\frac{2\pi}{3}j\right)\sigma_x+\sin\left(\frac{2\pi}{3}j\right)\sigma_y\right),\end{eqnarray}
where $1_n$ is an $n\times n$ unit matrix.
$\hat{H}_{BM}$ acts on its eigenfunctions
\begin{eqnarray}
\Psi_{n,\bk}(\br)=\sum_{\bg}\left(\begin{array}{c}a_{n,\bg}(\bk)\\ b_{n,\bg}(\bk)e^{i\bq_1\cdot\br}\end{array}\right)e^{i\bk\cdot\br}e^{i\bg\cdot\br},
\end{eqnarray}
where $\bg=m_1\bg_1+m_2\bg_2$ for integer $m_{1,2}$ and $\bg_{1,2}=\bq_{2,3}-\bq_1$.
The slow fields at the two valleys $\bK/\bK'$ are expanded in this `band' basis fermion annihilation operators $d_{\sigma,\bK/\bK',n,\bk}$ with crystal momentum $\bk$ in first moire Brillouin zone, and the band index $n$ as
\begin{eqnarray}\label{Eq:fields}
\chi_\sigma(\br)=\left(\begin{array}{cc} \psi_\sigma(\br)\\ \phi_\sigma(\br)\end{array}\right)
&=&\sum_{n\bk} \left(\begin{array}{cc}\Psi_{n,\bk}(\br)d_{\sigma,\bK,n,\bk}\\ \Psi^*_{n,\bk}(\br)d_{\sigma,\bK',n,-\bk-\bq_1}\end{array}\right).
\end{eqnarray}
It will be helpful for us to think about $H_{kin}$ as a lowest order gradient expansion of a continuum field theory\cite{Leon1}, with coupling constants that can flow due to $V_{int}$ under the $1^{st}$ stage of RG.

As pointed out in Ref.\onlinecite{Bernevig1}, if the small angle rotation in $\sigma_{\theta/2}$ is ignored, then $\hat{H}_{BM}$ enjoys a p-h symmetry for any value of $w_0$ and $w_1$,
\begin{eqnarray}
-i\mu_y \sigma_x\hat{H}^*_{BM} \sigma_x i\mu_y&=& -\hat{H}_{BM},
\end{eqnarray}
in that if $\Psi_{n,\bk}(\br)$ is an eigenstate of $\hat{H}_{BM}$ at $\bk$ with eigenvalue $\eps_{n,\bk}$, then
$-i\mu_y\sigma_x\Psi^*_{n,\bk}(\br)$ is an eigenstate at $-\bk-\bq_1$ with eigenvalue $-\eps_{n,\bk}$.
In what follows, we will neglect the small p-h asymmetric term which is two orders of magnitude smaller than $w_{0,1}$ and which we analyse in Ref\cite{SM}, and perform our RG assuming this approximate symmetry is present.

Up to an overall shift of the chemical potential, we can rewrite $V_{int}$ as
\begin{eqnarray}
\label{Eqn:Vint ph}
V_{int} &=& \frac{1}{2}\int d^2\br d^2\br' V(\br-\br') \delta\rho(\br)\delta\rho(\br'),\\
\delta\rho(\br)&=&\chi_\sigma^\dagger(\br)\chi_{\sigma}(\br)-\frac{1}{2}\{\chi_{\sigma}^\dagger(\br),\chi_{\sigma}(\br)\}.
\label{Eqn:delta rho}
\end{eqnarray}
For a pure Coulomb interaction $V(\br)=e^2/\eps r$. The Hamiltonian in Eqs.(\ref{Eqn:starting Hamiltonian kin})-(\ref{Eqn:starting Hamiltonian}) is defined at some high energy cut-off $\pm E_c$ which corresponds to a maximal value of the band index $n_c$ in our expansion. The parameters $v_F$, $w_0$, and $w_1$ should also be thought of as being fixed by a measurement at $E_c$.
The last term in (\ref{Eqn:delta rho}) is usually ignored, but for our RG, it will be helpful to express it as
\begin{eqnarray}
\frac{1}{2}\{\chi_{\sigma}^\dagger(\br),\chi_{\sigma}(\br)\}=\bar{\rho}_{E_c}(\br)&=&2\sum_{|\eps_{n\bk}|\leq E_c} \Psi^*_{n,\bk}(\br)\Psi_{n,\bk}(\br).
\end{eqnarray}
In the $1^{st}$ stage, we split $\chi_{\sigma}(\br)=\chi^{>}_{\sigma}(\br)+\chi^{<}_{\sigma}(\br)$ and integrate out the fast modes $\chi^{>}_{\sigma}(\br)$ with kinetic energy $E'_{c} < |\eps_{n,\bk}|\leq E_c$, such that $E'_c\gg w_{0,1}$. In this regime, the $V_{int}$ can be treated perturbatively. Its contribution to the slow mode Hamiltonian is then
\begin{eqnarray}\label{Eqn:Vint RG}
&&V_{int}\rightarrow\frac{1}{2}\int d^2\br d^2\br' V(\br-\br') \delta\rho^<(\br)\delta\rho^<(\br')\nonumber\\
&+&\frac{1}{2}\int d^2\br d^2\br' V(\br-\br'){\chi_\sigma^<}^\dagger(\br)\delta\mathcal{F}(\br,\br'){\chi^<_{\sigma}}(\br'),
\end{eqnarray}
where $\delta\rho^<(\br)={\chi^<_\sigma}^\dagger(\br)\chi^<_{\sigma}(\br)-\bar{\rho}_{E'_c}(\br)$ which follows from the p-h symmetry.
The correction to the $\hat{H}_{BM}$ comes from
\begin{eqnarray}
\delta\mathcal{F}(\br,\br')=\!\!\!\!\!\!\sum_{E'_c < |\eps_{n\bk}| \leq  E_c}\!\!\!\!\!\!\mbox{sign}(\eps_{n\bk})
\left(\begin{array}{cc}
f_{n,\bk}(\br,\br') & 0\\
0 & f^*_{n,\bk}(\br,\br')
\end{array}\right),
\end{eqnarray}
where $f_{n,\bk}(\br,\br')=\Psi_{n,\bk}(\br)\Psi^\dagger_{n,\bk}(\br')$.
We can now write
\begin{eqnarray}
\sum_{E'_c < |\eps_{n\bk}| \leq  E_c}\!\!\!\!\!\!\mbox{sign}(\eps_{n\bk})f_{n,\bk}(\br,\br')=\oint_\mathcal{C}\frac{dz}{2\pi i}\langle \br|\hat{G}(z)|\br'\rangle
\end{eqnarray}
where $\hat{G}(z)=\left(z-\hat{H}_{BM}\right)^{-1}$, and the contour $\mathcal{C}$ encloses the $z$-plane real line segment $\left(-E_c,-E'_c\right)$ in the clockwise, and segment $\left(E'_c,E_c\right)$ in the counterclockwise, sense.
As long as $E'_c\gg w_{0,1}$, the dominant contribution to the contour integral can be found by replacing
$\hat{G}(z)\approx \hat{G}_0(z)+\hat{G}_0(z)\hat{T}\hat{G}_0(z)+\mathcal{O}\left(\frac{w^2_{0,1}}{{E'}^2_c}\right).$
For small $E_c-E'_c$, we thus find that in the $1^{st}$ RG stage\cite{SM},
\begin{eqnarray}
\frac{dv_F}{d\ln E_c}&=&-\frac{e^2}{4\eps},\\
\frac{dw_0}{d\ln E_c}&=&0,\\
\frac{dw_1}{d\ln E_c}&=&-w_1\frac{e^2}{4\eps v_F},
\end{eqnarray}
and $e^2$, being the prefactor of a non-analytic term, does not renormalize when high energy modes are eliminated\cite{HerbutPRL2001}.
Integrating the above equations i.e. progressively reducing the cutoff to $E^*_c$ gives
\begin{eqnarray}\label{Eqn:RG magic angle}
\frac{w_1(E^*_c)}{v_F(E^*_c)}&=&\frac{w_1(E_c)}{v_F(E_c)},\\
\label{Eqn:chiral limit}
\frac{w_0(E^*_c)}{w_1(E^*_c)}&=&\frac{w_0(E_c)}{w_1(E_c)}\bigg/\left(1+\frac{e^2}{4\eps v_F(E_c)}\ln\frac{E_c}{E_c^*}\right).
\end{eqnarray}
The Eq.(\ref{Eqn:RG magic angle}) implies that the magic angle condition is largely insensitive to the renormalization.
The Eq.(\ref{Eqn:chiral limit}) shows that even if we start away from the chiral limit\cite{Grisha} at the UV scale $E_c$, at a lower energy scale $E^*_c$ we approach it.
Next, we combine this stage 1 RG with the non-perturbative (in moire potential) stage 2 numerical RG at $6w_1\gtrsim E^*_c$, but we stress that results are insensitive to the choice of  $E^*_c$ as long as $w_{1,0}/E^*_c$ is small so that stage 1 is under control.
The scaling collapse of the band structure shown in the Fig.(\ref{Fig:nonchiralscaling}) demonstrates this insensitivity for $w_1/v_Fk_\theta=0.5$, $e^2/v_F=2.2$, and $\eps=4.4$ with several choices of $n_c$. We also find an increase of the sublattice polarization and steepening of the Wilson loops along the RG evolution\cite{SM}, indicating a further approach of the chiral limit during the stage 2.

Note that at each step of our procedure we re-diagonalize the BM-like model in the subspace of the low energy bands corrected by $V_{int}$. We also re-express the $V_{int}$ in (\ref{Eqn:Vint ph}) in terms of the current (rotated) eigenstates of the BM model below the running energy cutoff, and because $\bar{\rho}_{E'_c}(\br)$ is invariant under the basis rotation, the p-h symmetry is explicitly preserved.
After the final step, we are thus left with two renormalized narrow bands per valley, and $V_{int}$ containing $\rho(\br)$ and $\bar{\rho}_0(\br)$ both expressed in terms of the final renormalized wavefunctions $\tilde{\Psi}_{n\pm,\bk}(\br)$, with the upper and lower bands denoted by $n+$ and $n-$, respectively.
Because the p-h symmetry is preserved during this procedure, we can choose $\tilde{\Psi}_{n-,\bk}(\br)=-i\mu_y\sigma_x \tilde{\Psi}^*_{n+,-\bk-\bq_1}(\br)$. Substitution of such field operators (\ref{Eq:fields}) gives
$\rho(\br)=\sum_{\bk\bk'}\sum_{\sigma=\uparrow,\downarrow}D^\dagger_{\bk\sigma}\mathcal{P}_{\bk\bk'}(\br) D_{\bk'\sigma}$, where within the narrow band $D^\dagger_{\bk\sigma}=(d^\dagger_{\bK,n+,\bk\sigma},d^\dagger_{\bK,n-,\bk\sigma},d^\dagger_{\bK',n+,\bk\sigma},d^\dagger_{\bK',n-,\bk\sigma})$. Suppressing $\bk\bk'$ and $\br$ dependence, $\mathcal{P}=b_0 1_4+b_1\tau_3\tilde{\sigma}_1+b_2 1_2\tilde{\sigma}_2 +b_3 \tau_3\tilde{\sigma}_3$, thus commuting with all 16 generators of spin-valley $U(4)$ symmetry\cite{Zaletel3} $1_4s_\mu$, $\tau_3 1_2s_\mu$, $\tau_2\tilde{\sigma}_2s_\mu$, $\tau_1\tilde{\sigma}_2s_\mu$, where $\mu=0,1,2,3$ and $\tau$ acts on valley, $\tilde{\sigma}$ on band, and $s$ on spin components ($s_0=1_2$).

If a state $|\Omega\rangle$ is annihilated by $\delta\rho(\br)$ for all $\br$, then it is a ground state at the strong coupling because $V_{int}$ is positive definite\cite{KangVafekPRL,Zaletel3}.
Moreover, any state obtained by a global $U(4)$ rotation is also a ground state, and, at the CNP, can be obtained from a fully filled valley polarized state\cite{KangVafekPRL,Zaletel3}. The exact n-body excitations above any one ground state can also be obtained by solving an $(n-1)$-body problem because
$V_{int}X|\Omega\rangle=\frac{1}{2}\int d^2\br d^2\br' V(\br-\br')
\left[\delta\rho(\br),\left[\delta\rho(\br'),X\right]\right]|\Omega\rangle$ and because the center of mass momentum is conserved.
Therefore, solving the operator eigen-equation
\begin{eqnarray}\label{Eq:eigenOp}
E X=\frac{1}{2}\int d^2\br d^2\br' V(\br-\br')
\left[\delta\rho(\br),\left[\delta\rho(\br'),X\right]\right],
\end{eqnarray}
provides the exact excitation states in the strong coupling limit. The Eq.(\ref{Eq:eigenOp}) can be readily solved for a single particle excitation and we show the result in the Ref.\onlinecite{SM}. Here we focus on the charge neutral excitations (excitons) $X=\sum_{mm'\bk}f^{\alpha\beta}_{mm'\bk}(\bq)d^\dagger_{\alpha m,\bk}d_{\beta m',(\bk-\bq)mod\bg}$, with spin/valley labels $\alpha,\beta$, by finding the eigenfunctions $f^{\alpha\beta}_{mm'\bk}(\bq)$. Due to the spin-valley $U(4)$ invariance of these equations, it is sufficient to solve for one spin and valley projection, the rest can be obtained by the symmetry.
The numerically obtained exciton spectrum at the magic angle is shown in the Fig.\ref{Fig:Exciton} for the center of mass momentum $\bq$ along the path shown in the Fig.~\ref{Fig:schematics}b. The quadratically vanishing dispersion of the lowest branch corresponds to the four $U(4)$ ferromagnetic Goldstone bosons\cite{Watanabe2020}. Under RG a second set of four modes softens. This corresponds to approaching the (``hidden'') $U(4)\times U(4)$ invariant chiral limit\cite{Zaletel3} with its $8$ Goldstone bosons. Their gap is a measure of the $U(4)\times U(4)$ anisotropy terms and for the parameters in the Fig.\ref{Fig:Exciton} this gap is $\Delta_{U(4)\times U(4)}\approx 0.2 e^2/\eps L_m\sim 5meV$; the gap vanishes at the chiral limit.
Note that the modes disperse despite a complete absence of kinetic energy terms due to the non-local structure of the projected density operators\cite{KangVafekPRL}.

The $H_{kin}$ breaks the spin-valley $U(4)$ symmetry down to $U(2)\times U(2)$ and causes splitting of the degenerate ground state manifold. We can obtain an upper bound on the resulting anisotropy terms from $2^{nd}$ order perturbation in (renormalized) kinetic energy (i.e. ``superexchange'') by replacing the energy of the excited states at $\fvec \Gamma$ with the lowest energy exciton that has a non-zero overlap on the kinetic energy operator ($E^{min}_{ph}\approx 2e^2/\eps L_m$ for Fig.\ref{Fig:Exciton}). For a spin independent valley rotation, parameterized by 3 Euler angles, $e^{\frac{i}{2}\alpha\tau_31_4}e^{\frac{i}{2}\omega\tau_2\sigma_2 1_2}e^{\frac{i}{2}\gamma\tau_31_4}$
we find that the energy splitting per unit cell, $\Delta_{U(4)}$, is bounded from above by $-\left(\sin^2\omega\right) 4\int d^2\bk \eps^2_{n+,\bk}/(A_{BZ}E^{min}_{ph})$. The
lowest energy state for such a rotation is the Kramers inter-valley coherent state\cite{Zaletel3} at $\omega=\frac{\pi}{2}$.
For the parameters in Fig.\ref{Fig:Exciton}, we find that $\Delta_{U(4)}<6.7 \times10^{-3}e^2/\eps L_m\sim 0.17 meV$, justifying the strong coupling approach.

The theory presented here can be extended to include the RPA effects and the p-h asymmetry, which will be important for any detailed quantitative comparison with experiments. Nevertheless, the Coulomb RG induced softening of the hidden symmetry collective modes, whose natural energy scale would normally be $\sim e^2/\eps L_m\sim 25meV$, suggests that they may not be frozen out even at $\sim 50K$.
Finally, our results offer a significant shift of the perspective in that the chiral limit\cite{Grisha} -- previously considered unphysical -- gains the status of an attractive mid-IR RG fixed point when $E_c/w_1\rightarrow \infty$.

We would like to thank Prof. B. Andrei Bernevig for
valuable discussions and for sharing their unpublished results
with us. O.~V.~is supported by NSF DMR-1916958 and partially by the National High Magnetic Field Laboratory through NSF Grant No.~DMR-1157490 and the State of Florida.
J.~K.~is supported by Priority Academic Program Development (PAPD) of Jiangsu Higher Education Institutions.

\appendix

\begin{widetext}

\newpage	
	
	\begin{center}
		\textbf{\large Supplemental Material for ``Towards the hidden symmetry in Coulomb interacting twisted bilayer graphene: renormalization group approach''}
	\end{center}

\setcounter{equation}{0}
\setcounter{figure}{0}
\setcounter{table}{0}
\makeatletter
\renewcommand{\thefigure}{S\arabic{figure}}

\section{Details of the 1$^{st}$ stage RG derivation for Coulomb interacting Bistritzer-MacDonal model}
For the contour $\mathcal{C}$ enclosing the $z$-plane real line segment $\left(-E_c,-E'_c\right)$ in the clockwise, and segment $\left(E'_c,E_c\right)$ in the counterclockwise, sense
\begin{eqnarray}
\oint_\mathcal{C} \frac{dz}{2\pi i}\langle \br |\hat{G}(z)|\br'\rangle&=&\int\frac{d^2\bk}{(2\pi)^2}\frac{d^2\bk'}{(2\pi)^2}
e^{i\bk\cdot\br} e^{-i\bk'\cdot\br'} \oint_\mathcal{C} \frac{dz}{2\pi i}\langle \bk |\hat{G}(z)|\bk'\rangle.
\end{eqnarray}
For $E'_c\gg w_{0,1}$, we can expand the Green's function to first non-trivial order in $\hat{T}$ and find
\begin{eqnarray}\label{EqSM:Gexpansion}
\oint_\mathcal{C} \frac{dz}{2\pi i}\langle \bk |\hat{G}(z)|\bk'\rangle &\approx&
\oint_\mathcal{C} \frac{dz}{2\pi i}\langle \bk |\hat{G}_0(z)|\bk'\rangle+
\oint_\mathcal{C} \frac{dz}{2\pi i}\langle \bk |\hat{G}_0(z)\hat{T}\hat{G}_0(z)|\bk'\rangle.
\end{eqnarray}
Without loss of generality, we can focus on one valley only, the contribution from the second one can be determined by time reversal symmetry.
The particle-hole symmetric BM Hamiltonian, acting on Bloch functions, is
\begin{eqnarray}
\hat{H}=\left(\begin{array}{cc}
v_F\sigma\cdot\bp & T(\br)e^{i\bq_1\cdot\br}\\
e^{-i\bq_1\cdot\br} T^\dagger(\br) & v_F\sigma\cdot(\bp+\bq_1)
\end{array}
\right)=\hat{H}_0+\hat{T},
\end{eqnarray}
and the interlayer tunneling term is
\begin{eqnarray}
T(\br)=\sum_{j=1}^3 T_j e^{-i\bq_j\cdot\br};\;
T_{j+1}=w_01+w_1\left(\begin{array}{cc} 0 & e^{-i\frac{2\pi}{3}j}\\
e^{i\frac{2\pi}{3}j} & 0\end{array}\right).
\end{eqnarray}
Here $\br$ and $\bp$ should be understood to be (first quantized) operators.
Therefore,
\begin{eqnarray}
\hat{G}_0(z)&=&\left(\begin{array}{cc}
\hat{g}_0(z,\bp) & 0\\
0 & \hat{g}_0(z,\bp+\bq_1)
\end{array}
\right)
\end{eqnarray}
where the intra-layer Green's function is
\begin{eqnarray}
\hat{g}_0(z,\bp)=\left(\omega-v_F\sigma\cdot\bp\right)^{-1}=\frac{1}{2}\sum_{s=\pm1}\frac{1+s\sigma\cdot\frac{\bp}{p}}{z-sv_Fp}
\end{eqnarray}
and
\begin{eqnarray}
\hat{G}_0(z)\hat{T}\hat{G}_0(z)&=&
\left(\begin{array}{cc}
0 & \hat{g}_0(z,\bp) T(\br)e^{i\bq_1\cdot\br}\hat{g}_0(z,\bp+\bq_1) \\
\hat{g}_0(z,\bp+\bq_1) e^{-i\bq_1\cdot\br} T^\dagger(\br)\hat{g}_0(z,\bp) & 0
\end{array}
\right).
\end{eqnarray}
Now,
\begin{eqnarray}
\langle \bk |\hat{g}_0(z,\bp)T(\br)e^{i\bq_1\cdot\br}\hat{g}_0(z,\bp+\bq_1)| \bk' \rangle &=&
\frac{1}{4}\sum_{ss'=\pm}\frac{1+s\sigma\cdot\frac{\bk}{k}}{z-sv_Fk}\langle \bk |T(\br)e^{i\bq_1\cdot\br}| \bk' \rangle
\frac{1+s'\sigma\cdot\frac{\bk'+\bq_1}{|\bk'+\bq_1|}}{z-s'v_F|\bk'+\bq_1|}\nonumber\\
&=&\frac{1}{4}\sum_{j=1}^3
\delta_{\bk',\bk+\bq_j-\bq_1}\sum_{ss'=\pm}\frac{1+s\sigma\cdot\frac{\bk}{k}}{z-sv_Fk}T_j
\frac{1+s'\sigma\cdot\frac{\bk+\bq_j}{|\bk+\bq_j|}}{z-s'v_F|\bk+\bq_j|},\label{EqSM:}
\end{eqnarray}
where we used
\begin{eqnarray}
\langle \bk |T(\br)e^{i\bq_1\cdot\br}| \bk' \rangle &=&
\sum_{j=1}^3T_j(2\pi)^2\delta\left(\bk'-(\bk+\bq_j-\bq_1)\right)\equiv
\sum_{j=1}^3T_j\delta_{\bk',\bk+\bq_j-\bq_1}.
\end{eqnarray}
Because it is novel, let us focus on the second term on the RHS in (\ref{EqSM:Gexpansion}).
Although formally the contribution to the contour integral comes from $s=s'$ and $s=-s'$, only the latter contributes to the RG flow. To see this, note that for $s=s'$ there is no contribution whatsoever if $v_Fk$ and $v_F|\bk+\bq_j|$ both lie inside, or both outside, the interval $\left(E'_c,E_c\right)$. There is a contribution only if one of them is outside of the interval. But, in that case, we can imagine extending the interval until both poles are included. This shows, that the contribution from adjacent shells cancels if $s=s'$.
Therefore, consider only $s=-s'$. Then, because $v_F|\bq_j|\ll E'_c$, we have
\begin{eqnarray}
&&\oint_{\mathcal{C}}\frac{dz}{2\pi i}
\left(\frac{1+\sigma\cdot\frac{\bk}{k}}{z-v_Fk}T_j
\frac{1-\sigma\cdot\frac{\bk+\bq_j}{|\bk+\bq_j|}}{z+v_F|\bk+\bq_j|}+\frac{1-\sigma\cdot\frac{\bk}{k}}{z+v_Fk}T_j
\frac{1+\sigma\cdot\frac{\bk+\bq_j}{|\bk+\bq_j|}}{z-v_F|\bk+\bq_j|}\right)\approx \nonumber\\
&&\frac{2}{v_Fk}\left(T_j-\sigma\cdot\frac{\bk}{k}T_j\sigma\cdot\frac{\bk}{k}\right)\Theta\left(E_c-v_Fk\right)\Theta\left(v_Fk-E'_c\right),
\end{eqnarray}
where $\Theta(x)$ is the Heaviside step function.
The component of $T_j$ proportional to $w_0$ is an identity matrix, which of course commutes through $\bk\cdot\sigma$. And because $\bk\cdot\sigma\bk\cdot\sigma=\bk^2 1$, there is no contribution to the renormalization of $w_0$.
So,
\begin{eqnarray}
&&\oint_\mathcal{C}\frac{dz}{2\pi i}\langle \bk |\hat{G}_0(z)\hat{T}\hat{G}_0(z)|\bk'\rangle \approx \nonumber\\
&&\sum_{j=1}^3\left(\begin{array}{cc}
0 & \delta_{\bk',\bk+\bq_j-\bq_1} \frac{\Theta\left(E_c-v_Fk\right)\Theta\left(v_Fk-E'_c\right)}{2v_Fk}\left(T_j-\sigma\cdot\frac{\bk}{k}T_j\sigma\cdot\frac{\bk}{k}\right)\\
\delta_{\bk,\bk'+\bq_j-\bq_1}\frac{\Theta\left(E_c-v_Fk'\right)\Theta\left(v_Fk'-E'_c\right)}{2v_Fk'}(T_j-\sigma\cdot\frac{\bk'}{k'}T_j\sigma\cdot\frac{\bk'}{k'}) & 0
\end{array}
\right).\nonumber\\
\end{eqnarray}
The contribution of this term to the correction to the moire tunneling potential is
\begin{eqnarray}
&&\frac{1}{2}\int d^2\br d^2\br' V(\br-\br')\int \frac{d^2\bk}{(2\pi)^2} \frac{d^2\bk'}{(2\pi)^2}e^{i\bk\cdot\br}e^{-i\bk'\cdot\br'}
{\psi^<_{\sigma}}^\dagger(\br)\left(\oint_\mathcal{C}\frac{dz}{2\pi i}\langle \bk |\hat{G}_0(z)\hat{T}\hat{G}_0(z)|\bk'\rangle\right) \psi^<_{\sigma}(\br')=\nonumber\\
&&\sum_{j=1}^3\int \frac{d^2\bq}{(2\pi)^2}
\left({\psi^<_{\sigma,\bq}}^\dagger\left(\begin{array}{cc}
0 & \Upsilon_j\left(\bq,E_c,E'_c\right)\\
0 & 0
\end{array}\right) \psi^<_{\sigma,\bq+\bq_j-\bq_1}+
{\psi^<}^\dagger_{\sigma,\bq+\bq_j-\bq_1}\left(\begin{array}{cc}
0 & 0\\
\Upsilon^\dagger_j\left(\bq,E_c,E'_c\right) & 0
\end{array}\right) \psi^<_{\sigma,\bq}\right).
\end{eqnarray}
where
\begin{eqnarray}
\Upsilon_j\left(\bq,E_c,E'_c\right)&=&\int\frac{d^2\bk}{(2\pi)^2}V_{\bk-\bq}
\frac{\Theta\left(E_c-v_Fk\right)\Theta\left(v_Fk-E'_c\right)}{4v_Fk}\left(T_j-\sigma\cdot\frac{\bk}{k}T_j\sigma\cdot\frac{\bk}{k}\right).
\end{eqnarray}
Because $v_Fq<E_c$, we can expand in powers of $v_Fq/E_c$.
Moreover, because
\begin{eqnarray}
\sigma\cdot\bk\sigma_{1,2}\sigma\cdot\bk&=&\pm\left(k^2_x-k^2_y\right)\sigma_{1,2}+2k_xk_y\sigma_{2,1},
\end{eqnarray}
the term $\sigma\cdot\bk T_j\sigma\cdot\bk$ will not contribute to the leading term in which $\bq$ is set to $0$ due to the angular integration.
For Coulomb interaction $V_\bk=2\pi e^2/(\eps k)$ we find
\begin{eqnarray}
\Upsilon_j\left(\bq,E_c,E'_c\right)&=&\frac{e^2}{4\eps v_F}T_{j,w_0=0} \ln\frac{E_c}{E'_c}+\ldots
\end{eqnarray}
where $\ldots$ are higher order terms in $v_Fq/E_c$.
Therefore, we find the RG equations for the interlayer couplings
\begin{eqnarray}
\frac{dw_0}{d\ln E_c}&=&0,\\
\frac{dw_1}{d\ln E_c}&=&-\frac{e^2}{4\eps v_F}w_1.
\end{eqnarray}
Clearly, as long as $v_Fq_j\ll E_c$, the expansion is in powers of $w_{0,1}/E_c$ and higher order terms in the expansion of BM Green's function, i.e. terms beyond $G_0TG_0$ will be suppressed by powers of $w_{0,1}/E_c$ and higher order gradients. The above term is the dominant correction to the BM interlayer tunneling as is consistent with the notion of the continuum model being a field theory expanded in powers of gradients\cite{Leon1}.

The contribution from the $G_0(\omega)$ term is standard and leads to
\begin{eqnarray}
\frac{dv_F}{d\ln E_c}&=&-\frac{e^2}{4\eps}.
\end{eqnarray}
Because $de^2/d\ln E_c=0$, the above equations are readily integrated. If we stop the renormalization at the scale $E^*_c\ll E_c$ we find
\begin{eqnarray}
w_1(E^*_c)&=&w_1(E_c)\left(1+\frac{e^2}{4\eps v_F(E_c)}\ln\frac{E_c}{E_c^*}\right)\\
v_F(E^*_c)&=&v_F(E_c)+\frac{e^2}{4\eps}\ln\frac{E_c}{E_c^*}\\
\Rightarrow \frac{w_1(E^*_c)}{v_F(E^*_c)}&=&\frac{w_1(E_c)}{v_F(E_c)}\\
\frac{w_0(E^*_c)}{w_1(E^*_c)}&=&\frac{w_0(E_c)}{w_1(E_c)}\frac{1}{\left(1+\frac{e^2}{4\eps v_F(E_c)}\ln\frac{E_c}{E_c^*}\right)}.
\end{eqnarray}
The above shows that even when we start away from the chiral limit\cite{Grisha} at the UV scale $E_c$, at a lower energy scale $E^*_c$ we approach it.
In practice, we find that, after we combine this stage 1 RG with the non-perturbative stage 2 numerical RG, our results are insensitive to the choice of $E^*_c$ as long as $w_1/E^*_c$ remains small.

\subsection{Weak particle-hole asymmetry}
The full BM Hamiltonian (without the small angle approximation), acting on Bloch functions, is
\begin{eqnarray}
\hat{H}=\left(\begin{array}{cc}
e^{-\frac{i}{4}\theta\sigma_3}v_F\sigma\cdot\bp e^{\frac{i}{4}\theta\sigma_3} & T(\br)e^{i\bq_1\cdot\br}\\
e^{-i\bq_1\cdot\br} T^\dagger(\br) & e^{\frac{i}{4}\theta\sigma_3} v_F\sigma\cdot(\bp+\bq_1)e^{-\frac{i}{4}\theta\sigma_3}
\end{array}
\right),
\end{eqnarray}
and the moire perturbation is
\begin{eqnarray}
T(\br)=\sum_{j=1}^3 T_j e^{-i\bq_j\cdot\br};\;
T_{j+1}=w_01+w_1\left(\begin{array}{cc} 0 & e^{-i\frac{2\pi}{3}j}\\
e^{i\frac{2\pi}{3}j} & 0\end{array}\right).
\end{eqnarray}

Now we perform a unitary transformation on $\hat{H}$ as
\begin{eqnarray}
\hat{H}&\rightarrow&
\left(\begin{array}{cc}
e^{\frac{i}{4}\theta\sigma_3} & 0\\
0 & e^{-\frac{i}{4}\theta\sigma_3}
\end{array}\right)
\left(\begin{array}{cc}
e^{-\frac{i}{4}\theta\sigma_3}v_F\sigma\cdot\bp e^{\frac{i}{4}\theta\sigma_3} & T(\br)e^{i\bq_1\cdot\br}\\
e^{-i\bq_1\cdot\br} T^\dagger(\br) & e^{\frac{i}{4}\theta\sigma_3} v_F\sigma\cdot(\bp+\bq_1)e^{-\frac{i}{4}\theta\sigma_3}
\end{array}
\right)
\left(\begin{array}{cc}
e^{-\frac{i}{4}\theta\sigma_3} & 0\\
0 & e^{\frac{i}{4}\theta\sigma_3}\end{array}\right)\\
&=&
\left(\begin{array}{cc}
v_F\sigma\cdot\bp & e^{\frac{i}{4}\theta\sigma_3}T(\br)e^{\frac{i}{4}\theta\sigma_3} e^{i\bq_1\cdot\br}\\
e^{-i\bq_1\cdot\br}e^{-\frac{i}{4}\theta\sigma_3}T^\dagger(\br)e^{-\frac{i}{4}\theta\sigma_3} &  v_F\sigma\cdot(\bp+\bq_1)
\end{array}
\right).
\end{eqnarray}
This means that
\begin{eqnarray}
T_{j+1}\rightarrow w_01 e^{\frac{i}{2}\theta\sigma_3}+w_1\left(\begin{array}{cc} 0 & e^{-i\frac{2\pi}{3}j}\\
e^{i\frac{2\pi}{3}j} & 0\end{array}\right)=
w_0\left(1\cos\frac{\theta}{2}+i\sigma_3\sin\frac{\theta}{2}\right)+w_1\left(\begin{array}{cc} 0 & e^{-i\frac{2\pi}{3}j}\\
e^{i\frac{2\pi}{3}j} & 0\end{array}\right),
\end{eqnarray}
and we see that we have another term proportional to $\sigma_3$; we denote it by $w_3$.
At the magic angle $\theta\approx 1.1^{\circ}$ we have $\sin\left(\frac{\theta}{2}\right)\approx 0.01$, therefore the ph symmetry breaking term $w_3$ starts two orders of magnitude smaller than the ph symmetric terms $w_0$ and $w_1$.

Note that the unitary transformation does not change the density and therefore does not change the Coulomb interaction.

The quantity which determined the renormalization is
$T_j-\frac{\bk}{k}\cdot\sigma T_j \frac{\bk}{k}\cdot\sigma$.
Because the ph asymmetric part of $T_j$ which is proportional to $\sigma_3$ anticommutes with $\bk\cdot\sigma$, its contribution to the RG flow has an extra factor of $2$ relative to the contribution from $w_1$.

We thus have
\begin{eqnarray}
\frac{dw_3}{d\ln E_c}&=&-2\frac{e^2}{4\eps v_F}w_3.
\end{eqnarray}
Integrating this equation gives
\begin{eqnarray}
w_3(E^*_c)&=&w_3(E_c)\left(1+\frac{e^2}{4\eps v_F(E_c)}\ln\frac{E_c}{E_c^*}\right)^2.
\end{eqnarray}
The ph asymmetry term $w_3$ thus grows under the RG, but because it starts out two order of magnitude smaller than the other terms, there is not enough dynamical range for it to become significant. For example, for $\eps=4.4$, $E_c=2eV$, $E^*_c=0.1eV$ we find that it increases from $0.01$ to $0.018$, which still makes it perturbatively small in the second stage RG. Even if the $\eps=1$, one would need an unrealistic $7$ orders of magnitude for $E_c/E^*_c$ to make this coupling of order unity.

\section{Sublattice polarization and Wilson loop evolution under RG; and the single particle dispersion in the strong coupling limit.}

\begin{figure}[h]
	\centering
	\subfigure[\label{Fig:bareSublatticePol}]{\includegraphics[width=0.49\columnwidth]{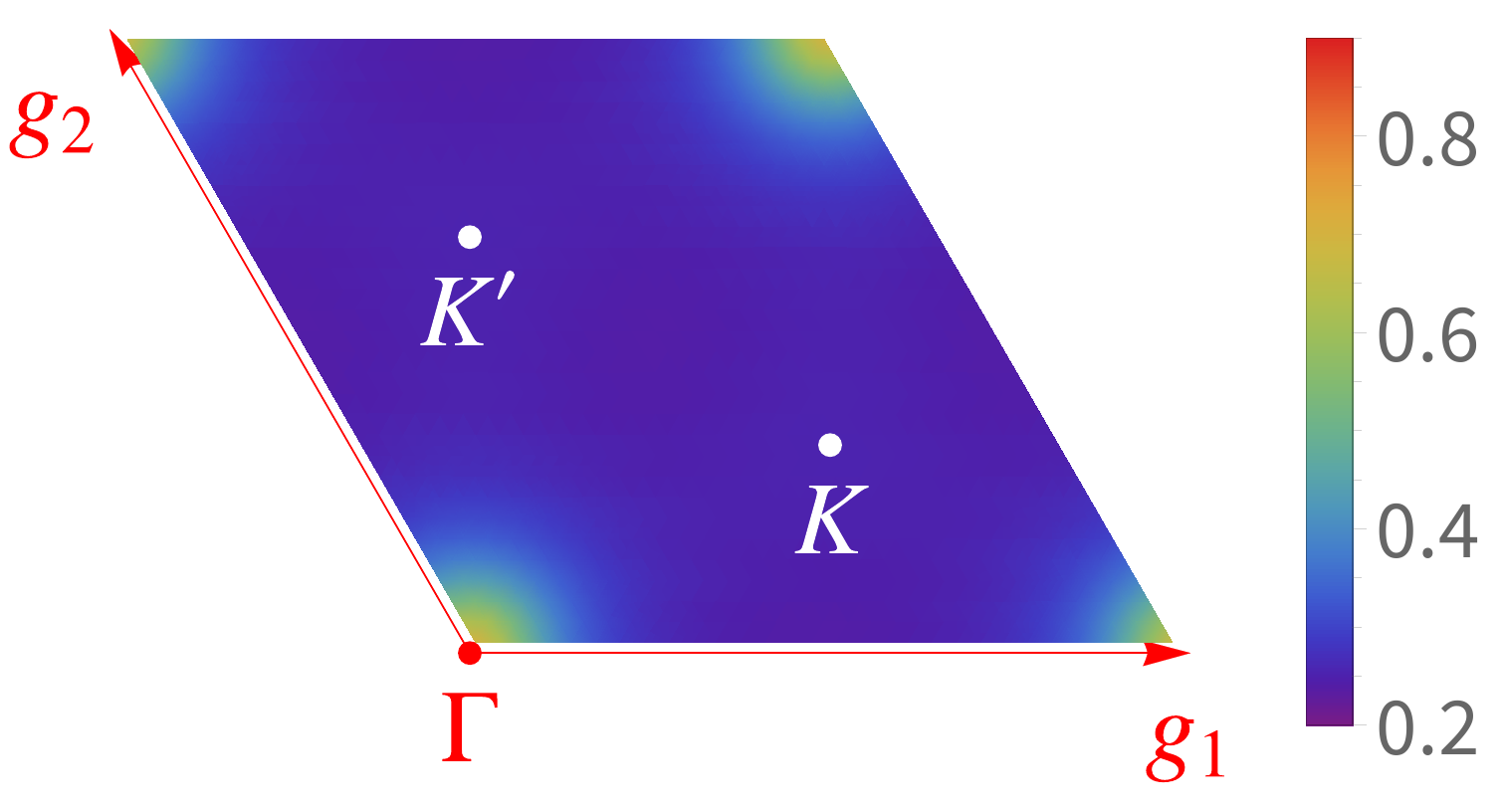}}
	\subfigure[\label{Fig:RenormalizedSublattice}]{\includegraphics[width=0.49\columnwidth]{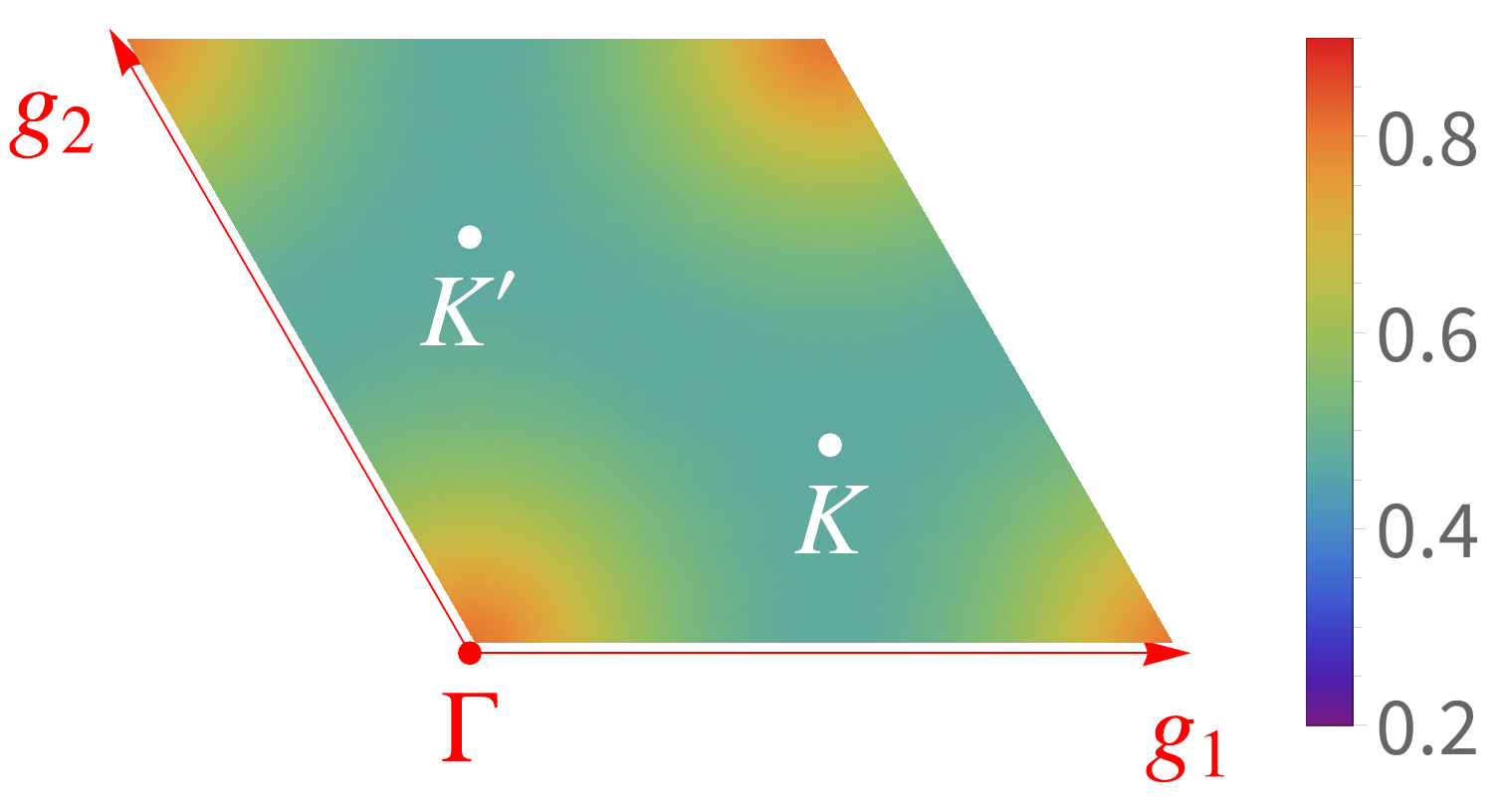}}
	\caption{The positive eigenvalue of the the sublattice polarization operator $1\sigma_3$ projected onto the two narrow bands at different $\bk$-points in BZ for un-renormalized BM model with $w_1/v_Fk_\theta=0.586$ (magic angle) and $w_0/w_1=0.83$ (a), and $w_0/w_1=0.83$ after 1$^{st}$ and $2^{nd}$ stage RG. The increase marks the approach of the chiral limit which is perfectly sublattice polarized i.e. the eigenvalue is $1$ for each $\bk$.}
	\label{Fig:sublattice}
\end{figure}
As explained in the e.g. Ref.\onlinecite{KangVafekPRB} the eigenstates of the projected operator
\begin{eqnarray}
\hat O = \hat P e^{-i \frac1N_1 \bg_1 \cdot  {\bf r}} \hat P,
\end{eqnarray}
are hybrid Wannier states. Here $\hat P$ is the projection operator onto the narrow bands. $\bg_1$ is the primitive vector of the reciprocal lattice shown in the Fig 1b of the main text, and $N_1$ is the number of unit cells along the direction of $\bL_1$ in the entire lattice with periodic boundary conditions.

We thus have
\begin{eqnarray}
\hat O |w_{\pm}(n, k \bg_2) \rangle = e^{-2\pi i\frac1N_1\left(n+\langle x_\pm\rangle_k/L_m\right)}|w_{\pm}(n, k \bg_2) \rangle  .
\end{eqnarray}
The hybrid WSs $|w_{\alpha}(n, k \bg_2) \rangle$ are labeled by their momentum $k$ along $\bg_2$ which is conserved by $\hat O$ and the index $n$ of the unit cell along $\bL_1$; $\alpha=\pm 1$ labels their winding number. Unlike the familiar lowest Landau level wavefunctions in the Landau gauge, the shapes of our hybrid WSs for the narrow bands depend on the momentum index $k$.

The $\langle x_\pm \rangle$ physically represents the average of the position operator within each 1D unit cell whose dependence on the conserved momentum $k$ is shown in the Fig.~\ref{Fig:Wilson} for various stages of renormalization. The two curves display the winding numbers of $\pm 1$ as the momentum $\bk$ increases from 0 to $\bg_2$
i.e. the average position of one set of states slides to the right and the other set of states to the left under the increase of the wavenumber $k$, similar to Landau gauge Landau level states in opposite magnetic field. The monotonic steepening of these curves under RG marks the approach of the chiral limit (see also Fig 3 of Ref.\onlinecite{KangVafekPRB}).

\begin{figure}[h]
	\centering
	\includegraphics[width=0.8\columnwidth]{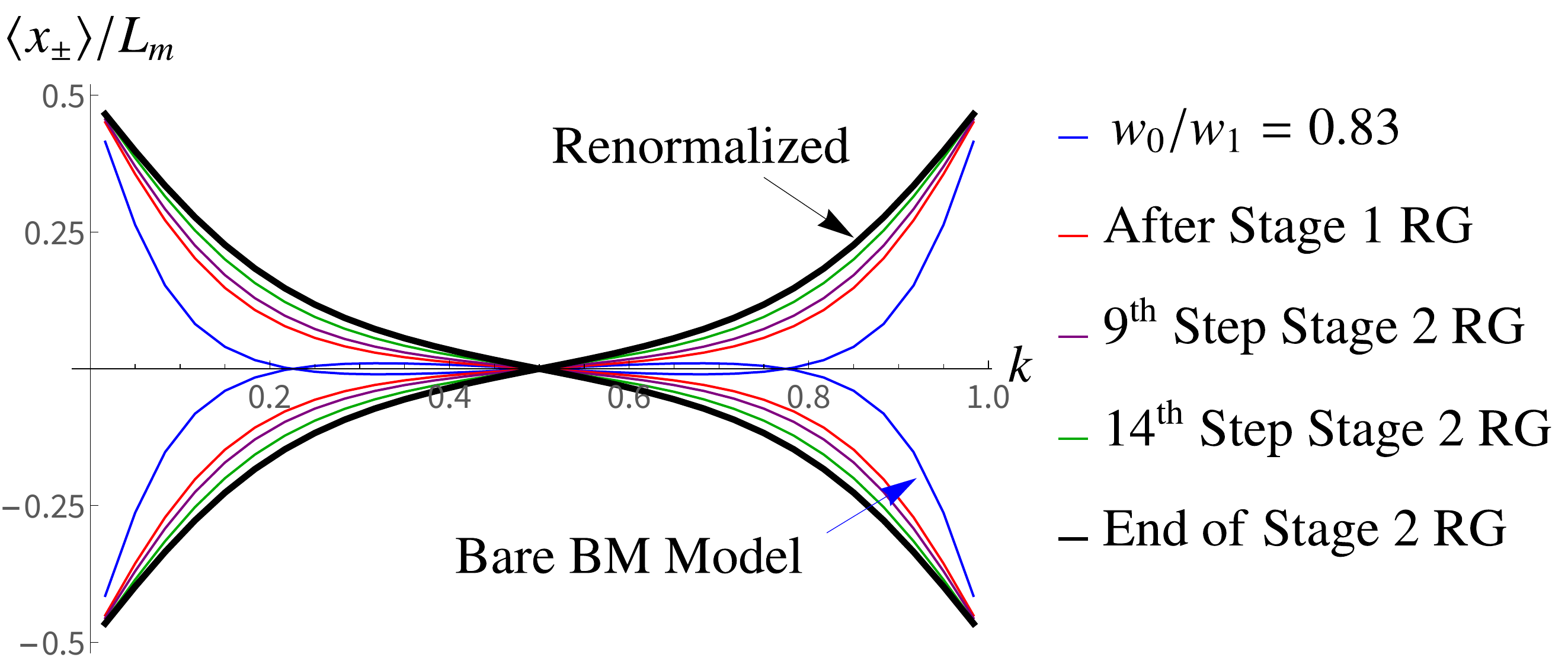}
	\caption{The RG evolution of the Wilson loop eigenvalues for the two narrow bands in valley $\bK$ for the unrenormalized narrow bands with $w_0/w_1=0.83$ (blue), renormalized after stage 1 (red), after stage 1 and $9^{th}$ step of stage 2 RG (purple), after stage 1 and $14^{th}$ step of stage 2 RG (green) and after both stage 1 and stage 2 (black). The steepening of the Wilson loop marks the approach of the chiral limit. The parameters for the RG are the same as the Fig 3 in the main text.}
	\label{Fig:Wilson}
\end{figure}
\begin{figure}[h]
	\centering
	\subfigure[\label{Fig:singleParticle:Disprsion}]{\includegraphics[width=0.49\columnwidth]{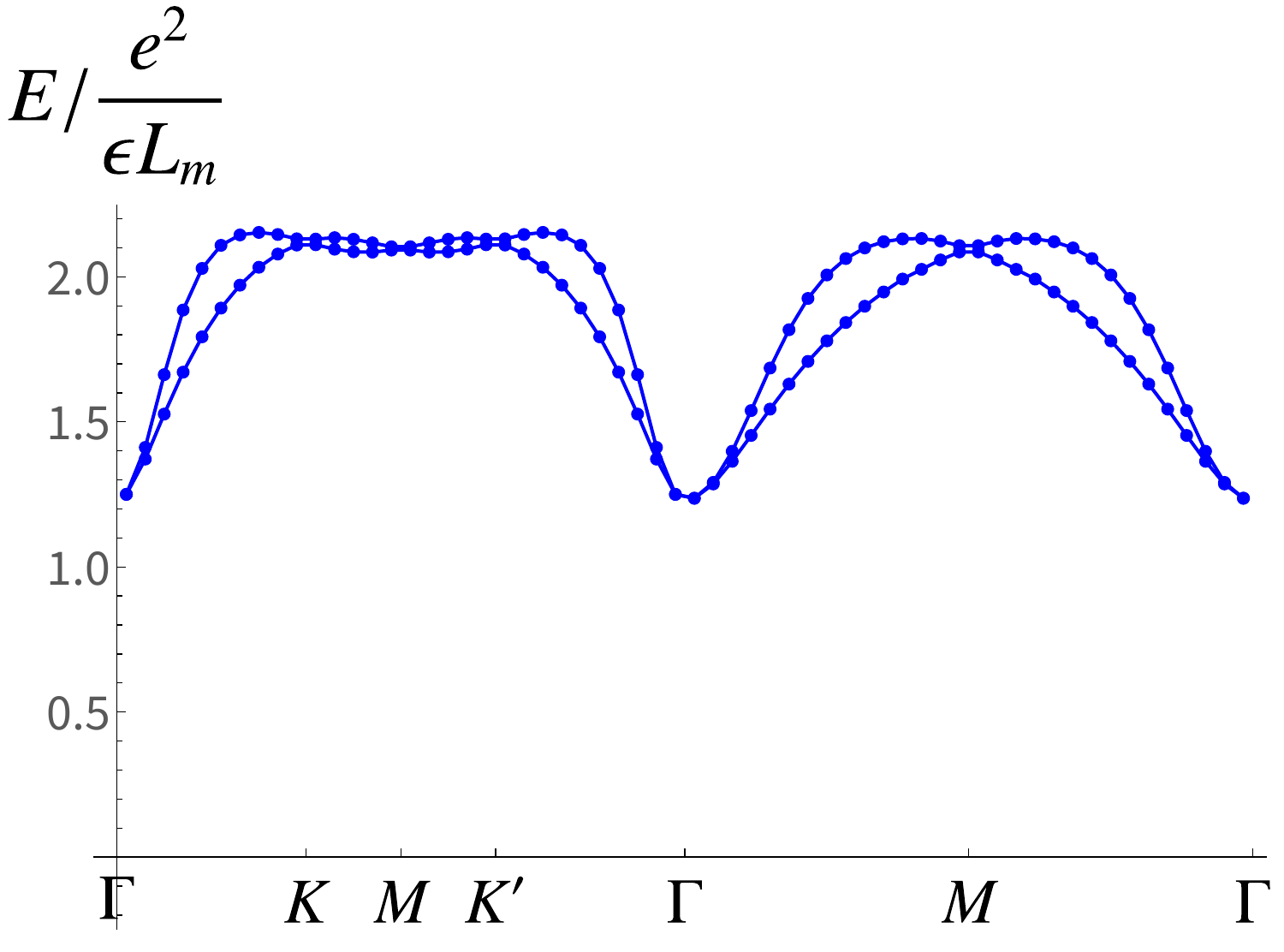}}
	\subfigure[\label{Fig:singleParticle:Filling}]{\includegraphics[width=0.49\columnwidth]{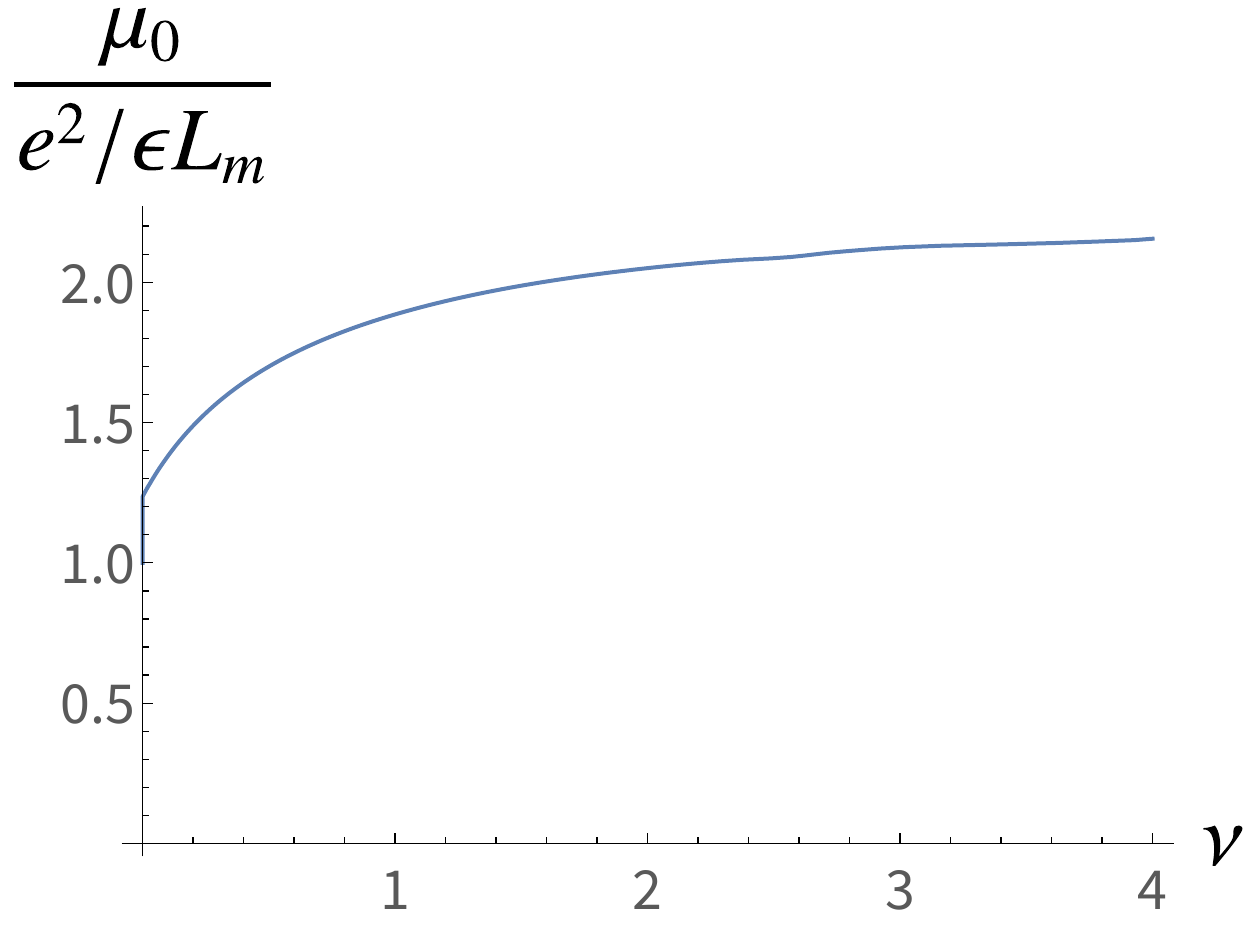}}
	\caption{ (a) The single particle dispersion in the strong coupling limit for $w_0/w_1=0.83$ and $w_1/v_Fk_\theta=0.586$ after the $1^{st}$ and $2^{nd}$ stage RG. The BZ cut is shown in the main text Fig 1b. (b) The chemical potential $\mu_0$ as the function of the filling factor $\nu$ assuming the single particle excitations are non-interacting.}
	\label{Fig:singleParticle}
\end{figure}

The single particle dispersion obtained in the strong coupling limit described in the main text is shown in the Fig.\ref{Fig:singleParticle}a. Note that unlike in a Hubbard model, despite being in the strong coupling, the dispersion is not flat, and may give rise to Fermi pressure for a finite density of single particle excitations, as illustrated in the Fig.\ref{Fig:singleParticle}b under the assumption that the excitations are non-interacting.
The single particle dispersion completely determines the onset of the two-particle continuum shown by the red curve in the Fig.3 of the main text.

\section{Outline of the numerical recipe for the stage 2 RG}

At the start of the $2^{nd}$ stage, we have $n_c$ bands above and $n_c$ bands below the CNP at each valley and for each spin projection.
As discussed in the main text, the (self-energy) correction to the one body part of the Hamiltonian is
\begin{eqnarray}
&&\frac{1}{2}\int d^2\br d^2\br' V(\br-\br'){\chi_\sigma^<}^\dagger(\br)\delta\mathcal{F}(\br,\br'){\chi^<_{\sigma}}(\br'),
\end{eqnarray}
where $\delta\mathcal{F}(\br,\br')$ comes only from the top-most ($n_c$) and the bottom-most ($-n_c$) bands
\begin{eqnarray}
\delta\mathcal{F}(\br,\br')=\sum_{\bk}\sum_{s=\pm 1}
\left(\begin{array}{cc}
s f_{sn_c,\bk}(\br,\br') & 0\\
0 & s f^*_{sn_c,\bk}(\br,\br')
\end{array}\right),
\end{eqnarray}
and where $f_{n,\bk}(\br,\br')=\Psi_{n,\bk}(\br)\Psi^\dagger_{n,\bk}(\br')$.

Without loss of generality we can focus on the valley ${\bf K}$, where the slow modes are
\begin{eqnarray}
\psi^<_\sigma(\br')&=&\sum_{\bk'}\sum_{|n|<n_c} \Psi_{n,\bk'}(\br')d_{\sigma,\bK,n,\bk'}.
\end{eqnarray}
The self-energy correction is then
\begin{eqnarray}
&&
\sum_{\bk''}\sum_{s=\pm}\frac{s}{2}\int d^2\br d^2\br' V(\br-\br'){\psi_\sigma^<}^\dagger(\br)
\Psi_{s n_c,\bk''}(\br)\Psi^\dagger_{sn_c,\bk''}(\br')
{\psi^<_{\sigma}}(\br')\\
&=&
\sum_{\bk,\bk'}\sum_{|n|<n_c}\sum_{|n'|<n_c}
d^\dagger_{\sigma,\bK,n,\bk}d_{\sigma,\bK,n',\bk'}
\sum_{\bk''}\sum_{s=\pm}\frac{s}{2}\int d^2\br d^2\br' V(\br-\br')\Psi^\dagger_{n,\bk}(\br)
\Psi_{s n_c,\bk''}(\br)\Psi^\dagger_{sn_c,\bk''}(\br')
\Psi_{n',\bk'}(\br').
\end{eqnarray}
For the Coulomb interaction,
\begin{eqnarray}
V(\br)&=&\frac{e^2}{\eps r}=\int \frac{d^2\bq}{(2\pi)^2}\frac{2\pi e^2}{\eps q}e^{i\bq\cdot\br} = \frac{1}{N_{uc}A_{uc}}\sum_{\bq}\frac{2\pi e^2}{\eps q}e^{i\bq\cdot\br},
\end{eqnarray}
where $N_{uc}$ is the number of moire unit cells and $A_{uc}=\frac{4\pi^2}{(\hat{z}\times\bg_1)\cdot\bg_2}=\frac{8\pi^2}{3\sqrt{3}k^2_\theta}$ is the area of the moire unit cell; $\bq_1=k_\theta(0,-1)$, $\bq_{2,3}=k_\theta\left(\pm\frac{\sqrt{3}}{2},\frac{1}{2}\right)$; $\bg_{1,2}=\bq_{2,3}-\bq_1$.
Substituting the above and using the Bloch periodicity of the wavefunctions, we find that
\begin{eqnarray}
&&\int d^2\br d^2\br' V(\br-\br')\Psi^\dagger_{n,\bk}(\br)
\Psi_{s n_c,\bk''}(\br)\Psi^\dagger_{sn_c,\bk''}(\br')
\Psi_{n',\bk'}(\br')=\\
&&\frac{1}{N_{uc}A_{uc}}\sum_\bq\frac{2\pi e^2}{\eps q}
\sum_{\bR,\bR'}e^{i(\bk''-\bk+\bq)\cdot\bR}e^{i(\bk'-\bk''-\bq)\cdot\bR'}
\int_{uc} d^2\br d^2\br' e^{i\bq\cdot(\br-\br')}\Psi^\dagger_{n,\bk}(\br)
\Psi_{s n_c,\bk''}(\br)\Psi^\dagger_{sn_c,\bk''}(\br')
\Psi_{n',\bk'}(\br')\\
&=&
\frac{N_{uc}}{A_{uc}}\sum_{\bg_\bq}\frac{2\pi e^2}{\eps |\bk-\bk''+\bg_\bq|}
\delta_{\bk,\bk'}
\int_{uc} d^2\br e^{i(\bk-\bk''+\bg_\bq)\cdot\br}\Psi^\dagger_{n,\bk}(\br)
\Psi_{s n_c,\bk''}(\br)
\int_{uc} d^2\br' e^{-i(\bk-\bk''+\bg_\bq)\cdot\br'}\Psi^\dagger_{sn_c,\bk''}(\br')
\Psi_{n',\bk}(\br'),
\end{eqnarray}
where we wrote $\bq=\bk_\bq+\bg_\bq$ and divided it into $\bk_\bq$ that contains the fractional part of $\bq$ and $\bg_\bq$ which is the integer multiple of reciprocal lattice unit vectors, and performed the Bravais $\bR$ lattice sums. The $\br$ and $\br'$ integrals are over the moire unit cell.

We thus have
\begin{eqnarray}
\sum_{\bk''}\sum_{s=\pm}\frac{s}{2}\int d^2\br d^2\br' V(\br-\br'){\psi_\sigma^<}^\dagger(\br)
\Psi_{s n_c,\bk''}(\br)\Psi^\dagger_{sn_c,\bk''}(\br')
{\psi^<_{\sigma}}(\br')=
\sum_{\bk}\sum_{|n|<n_c}\sum_{|n'|<n_c}
d^\dagger_{\sigma,\bK,n,\bk}\Sigma_{n,n'}(\bK,\bk,n_c)d_{\sigma,\bK,n',\bk},\nonumber\\
\end{eqnarray}
where
\begin{eqnarray}
&&\Sigma_{n,n'}(\bK,\bk,n_c)=\nonumber\\
&&\sum_{s=\pm}\frac{s}{2}\frac{N_{uc}}{A_{uc}}\sum_{\bk''}\sum_{\bg_\bq}\frac{2\pi e^2}{\eps |\bk-\bk''+\bg_\bq|}
\int_{uc} d^2\br e^{i(\bk-\bk''+\bg_\bq)\cdot\br}\Psi^\dagger_{n,\bk}(\br)
\Psi_{s n_c,\bk''}(\br)
\int_{uc} d^2\br' e^{-i(\bk-\bk''+\bg_\bq)\cdot\br'}\Psi^\dagger_{sn_c,\bk''}(\br')
\Psi_{n',\bk}(\br').\nonumber\\
\end{eqnarray}
The overlap integrals in $\Sigma_{n,n'}(\bK,\bk,n_c)$ are readily performed from the numerical diagonalization of the BM model in momentum space.

After eliminating the highest and the lowest bands, our renormalized Hamiltonian is then
\begin{eqnarray}
&&\sum_{\bk}\sum_{|n|<n_c}\sum_{|n'|<n_c}
d^\dagger_{\sigma,\bK,n,\bk}\left(\delta_{n,n'}\eps_{n,\bk}+\Sigma_{n,n'}(\bK,\bk,n_c)\right)d_{\sigma,\bK,n',\bk},\nonumber\\
&+&\sum_{\bk}\sum_{|n|<n_c}\sum_{|n'|<n_c}
d^\dagger_{\sigma,\bK',n,\bk}\left(\delta_{n,n'}\eps_{n,-\bk-\bq_1}+\Sigma_{n,n'}(\bK',\bk,n_c)\right)d_{\sigma,\bK',n',\bk},\nonumber\\
&+&
\frac{1}{2}\int d^2\br d^2\br' V(\br-\br') \delta\rho^<(\br)\delta\rho^<(\br'),
\end{eqnarray}
where
\begin{eqnarray}\label{EqSM:delta rho less}
\delta\rho^<(\br)={\chi^<_\sigma}^\dagger(\br)\chi^<_{\sigma}(\br)-2\sum_{\bk}\sum_{|n|<n_c} \Psi^*_{n,\bk}(\br)\Psi_{n,\bk}(\br),
\end{eqnarray}
and where the modes $\chi^<_{\sigma}(\br)$ are composed of eigenstates whose band indices range from $-n_c+1$ to $n_c-1$.

In the next step, we diagonalize $\delta_{n,n'}\eps_{n,\bk}+\Sigma_{n,n'}(\bK,\bk,n_c)$ and similarly in the valley $\bK'$.
We then use the diagonalizing unitary transformation to re-express the interaction energy in terms of the new $d$-operators, thus rotating the remaining eigenfunctions, $\Psi_{n,\bk}\rightarrow \tilde\Psi_{n,\bk}=\mathcal{U}_{n,n'}(\bk)\Psi_{n',\bk}$, for the remaining $2n_c-2$ bands. The last term in (\ref{EqSM:delta rho less}) is invariant because the transformation is unitary and because the $n$-sum involves all modes that are being mixed by $\mathcal{U}$. Thus
$\sum_{|n|<n_c} \Psi^*_{n,\bk}(\br)\Psi_{n,\bk}(\br)= \sum_{|n|<n_c} \tilde\Psi^*_{n,\bk}(\br)\tilde\Psi_{n,\bk}(\br)$.

This completes one step of the numerical RG, which reduces the $n_c$ by $1$, and which we iterate until we reach the narrow bands, i.e. until $\Sigma_{n,n'}$ is just a $2\times 2$ matrix.

\end{widetext}


\begin{thebibliography}{90}
	
\bibitem{GeimNovoselovNatPhys2011} D.C. Elias et al., ``Dirac cones reshaped by interaction effects in
suspended graphene'', Nat.Phys. \textbf{7} 701 (2011)

\bibitem{GeimNovoselovPNAS2013} G.L. Yu et al., ``Interaction phenomena in graphene seen through quantum capacitance'', PNAS \textbf{110}, 3282 (2013).

\bibitem{Gonzalez1994} J. Gonzalez, F. Guinea. and A.H. Vozmediano, ``Non-Fermi liquid behavior of electrons in the half-filled honeycomb lattice (A renormalization group approach'', Nucl. Phys. B
\textbf{424} 595, (1994).

\bibitem{VafekPRL2007} O. Vafek, ``Anomalous Thermodynamics of Coulomb-Interacting Massless Dirac Fermions in Two Spatial Dimensions'' Phys. Rev. Lett. \textbf{98}, 216401 (2007)

\bibitem{SheehyPRL2007} D.E. Sheehy and J. Schmalian, ``Quantum Critical Scaling in Graphene'',
Phys. Rev. Lett. \textbf{99}, 226803 (2007).

\bibitem{BorghiSSC2009} G. Borghi, M. Polini, R. Asgari, and A.H. MacDonald, ``Fermi velocity enhancement in monolayer and bilayer graphene'', Solid State Comm. \textbf{149} 1117 (2009).

\bibitem{BarnesPRB2014} E. Barnes, E. H. Hwang, R. E. Throckmorton, and S. Das Sarma, ``Effective field theory, three-loop perturbative expansion, and their experimental implications in graphene many-body effects''
Phys. Rev. B \textbf{89}, 235431 (2014).

\bibitem{BMModel} R. Bistritzer and A. H. MacDonald, ``Moire bands in twisted double-layer graphene,'' Proc. Natl. Acad. Sci. U.S.A.\textbf{108}, 12233 (2011).

\bibitem{Pablo1} Y. Cao, V. Fatemi, A. Demir, S. Fang, S. L. Tomarken, J. Y. Luo, J. D. Sanchez-Yamagishi, K. Watanabe, T. Taniguchi, E. Kaxiras, R. C. Ashoori, and P. Jarillo-Herrero, ``Correlated insulator behaviour at half-filling in magic-angle graphene superlattices,'' Nature \textbf{556}, 43 (2018).

\bibitem{Pablo2} Y. Cao, V. Fatemi, S. Fang, K. Watanabe, T. Taniguchi, E. Kaxiras, and P. Jarillo-Herrero, ``Unconventional superconductivity in magic-angle graphene superlattices,'' Nature \textbf{556}, 80 (2018).
	
\bibitem{Cory1} M. Yankowitz, S. Chen, H. Polshyn, Y. Zhang, K. Watanabe, T. Taniguchi, D. Graf, A. F. Young, and C. R. Dean, ``Tuning superconductivity in	twisted bilayer graphene,'', Science \textbf{363}, 1059 (2019).
	
\bibitem{David} A. L. Sharpe, E. J. Fox, A. W. Barnard, J. Finney, K. Watanabe, T. Taniguchi, M. A. Kastner, and D. Goldhaber-Gordon, ``Emergent ferromagnetism near	three-quarters filling in twisted bilayer graphene,'' Science \textbf{365}, 605 (2019).

\bibitem{Young} M. Serlin, C. L. Tschirhart, H. Polshyn, Y. Zhang, J. Zhu, K. Watanabe, T. Taniguchi, L. Balents, and A. F. Young, ``Intrinsic quantized anomalous hall effect in a moire heterostructure,'' Science science.aay5533 (2019).
	
\bibitem{Cory2} A. Kerelsky, L. J McGilly, D. M. Kennes, L. Xian, M. Yankowitz, S. Chen, K. Watanabe, T. Taniguchi, J. Hone, C. Dean, et al., ``Maximized electron interactions at the magic angle in twisted bilayer graphene'', Nature \textbf{572}, 95 (2019).
	
\bibitem{Dmitry1} X. Lu, P. Stepanov, W. Yang, M. Xie, M. A. Aamir, I. Das, C. Urgell, K. Watanabe, T. Taniguchi, G. Zhang, A. Bachtold, A. H. MacDonald, and D. K. Efetov, ``Superconductors, orbital magnets and correlated states in magic-angle bilayer graphene,'' Nature \textbf{574}, 653 (2019).
	
\bibitem{Yazdani} Y. Xie, B. Lian, B. Jack, X. Liu, C.-L. Chiu, K. Watanabe, T. Taniguchi, B. A. Bernevig, and A. Yazdani, ``Spectroscopic signatures of many body correlations in magic-angle twisted bilayer graphene,'' Nature \textbf{572}, 101 (2019).
	
\bibitem{Ashoori} S. L. Tomarken, Y. Cao, A. Demir, K. Watanabe, T. Taniguchi, P. Jarillo-Herrero, and R. C. Ashoori, ``Electronic compressibility of magic-angle graphene superlattices,'' Phys. Rev. Lett. \textbf{123}, 046601 (2019).
	
\bibitem{Eva} Y. Jiang, X. Lai, K. Watanabe, T. Taniguchi, K. Haule, J. Mao, and E. Y. Andrei, ''Charge order and broken rotational symmetry in magic-angle	twisted bilayer graphene,'' Nature \textbf{573}, 91 (2019).	

\bibitem{Stevan} Y. Choi, J. Kemmer, Y. Peng, A. Thomson, H. Arora, R. Polski, Y. Zhang, H. Ren, J. Alicea, G. Refael, F. von Oppen, K. Watanabe, T. Taniguchi, and S. Nadj-Perge, ``Electronic correlations in twisted bilayer graphene near the magic angle'', Nat. Phys. \textbf{15}, 1174 (2019).


\bibitem{Dmitry2} P. Stepanov, I. Das, X. Lu, A. Fahimniya, K. Watanabe, T. Taniguchi, F. H. L. Koppens, J. Lischner, L. Levitov, D. K. Efetov, ``Untying the insulating and superconducting orders in magic-angle graphene'', Nature \textbf{583}, 375 (2020).
	
\bibitem{Yazdani2} D. Wong, K. P. Nuckolls, M. Oh, B. Lian, Y. Xie, S. Jeon, K. Watanabe, T. Taniguchi, B. A. Bernevig, and A. Yazdani, ``Cascade of electronic transitions in magic-angle twisted bilayer graphene,'' Nature \textbf{582}, 198 (2020).
	
	
\bibitem{Shahal} U. Zondiner, A. Rozen, D. Rodan-Legrain, Y. Cao, R. Queiroz, T. Taniguchi, K. Watanabe, Y. Oreg, F. von Oppen, A. Stern, E. Berg, P. Jarillo-Herrero, and S. Ilani, ``Cascade of Phase Transitions and Dirac Revivals in Magic Angle Graphene,'' Nature \textbf{582}, 203 (2020).
	
\bibitem{Young2} Y. Saito, J. Ge, K. Watanabe, T. Taniguchi, and A. F. Young, ``Independent superconductors and correlated insulators in twisted bilayer graphene,'' Nat. Phys. \textbf{16}, 926 (2020).

\bibitem{YuanCao2020} Y. Cao, D. Rodan-Legrain, J. M. Park, F. N. Yuan, K. Watanabe, T. Taniguchi, R. M. Fernandes, L. Fu, and P. Jarillo-Herrero, ``Nematicity and Competing Orders in Superconducting Magic-Angle Graphene,'' arXiv:2004.04148.

\bibitem{Young3} C. L. Tschirhart, M. Serlin, H. Polshyn, A. Shragai, Z. Xia, J. Zhu, Y. Zhang, K. Watanabe, T. Taniguchi, M. E. Huber, A. F. Young, ``Imaging orbital ferromagnetism in a moire Chern insulator'', arXiv:2006.08053.

\bibitem{Xu} C. Xu and L. Balents, ``Topological Superconductivity in Twisted Multilayer Graphene'', Phys. Rev. Lett. \textbf{121}, 087001 (2018).
		
\bibitem{KangVafekPRX} J.~Kang and O.~Vafek, ``Symmetry, maximally localized Wannier states, and a low-energy model for twisted bilayer graphene narrow bands,'' Phys. Rev. X \textbf{8}, 031088 (2018).

\bibitem{LiangPRX1} M.~Koshino, N. F. Q. Yuan, T.~Koretsune, M.~Ochi, K.~Kuroki, and L.~Fu, ``Maximally localized wannier orbitals and the extended hubbard model for twisted bilayer graphene,'' Phys. Rev. X \textbf{8}, 031087 (2018).
	
\bibitem{Senthil1} H. C. Po, L. Zou, A. Vishwanath, and T. Senthil, ``Origin of mott insulating behavior and superconductivity in twisted bilayer graphene,'' Phys. Rev. X {\bf 8}, 031089 (2018).
	
\bibitem{FanYang} C.-C. Liu, L.-D. Zhang, W.-Q. Chen, and F. Yang, ``Chiral spin density wave and d + id superconductivity in the magic-angle-twisted bilayer graphene'', Phys. Rev. Lett. \textbf{121}, 217001 (2018).
	
\bibitem{FengchengSC} F. Wu, A. H. MacDonald, and I. Martin, ``Theory of phonon-mediated superconductivity in twisted bilayer graphene'', Phys. Rev. Lett. \textbf{121}, 257001 (2018).

\bibitem{LiangPRX2} Hiroki Isobe, Noah F. Q. Yuan, and Liang Fu, ``Unconventional superconductivity and density waves in twisted bilayer
	graphene,'' Phys. Rev. X \textbf{8}, 041041 (2018).
	
\bibitem{Guo} H.~Guo, X.~Zhu, S.~Feng, and R. T. Scalettar, ``Pairing symmetry of interacting fermions on a twisted bilayer graphene superlattice'', Phys. Rev. B \textbf{97}, 235453 (2018).
	
	
\bibitem{GuineaPNAS} F. Guinea and N. R Walet, ``Electrostatic effects, band distortions, and superconductivity in twisted graphene bilayers,'' Proc. Natl. Acad. Sci. U.S.A. \textbf{115}, 13174 (2018).

\bibitem{Thomson} A. Thomson, S. Chatterjee, S. Sachdev, and M. S.Scheurer, ``Triangular antiferromagnetism on the honeycomb lattice of twisted bilayer graphene'', Phys. Rev. B \textbf{98}, 075109 (2018).

\bibitem{Kivelson} J. F. Dodaro, S. A. Kivelson, Y. Schattner, X. Q. Sun, and C. Wang, ``Phases of a phenomenological model of twisted bilayer graphene,'' Phys. Rev. B \textbf{98}, 075154 (2018).


\bibitem{SenthilTop} L. Zou, H. C. Po, A. Vishwanath, and T. Senthil, ``Band structure of twisted bilayer graphene: Emergent symmetries, commensurate approximants, and Wannier obstructions,''  Phys. Rev. B {\bf 98}, 085435 (2018).

\bibitem{Kuroki} M.~Ochi, M.~Koshino, K.~Kuroki, ``Possible correlated insulating states in magic-angle twisted bilayer graphene under strongly competing interactions'', Phys. Rev. B {\bf 98}, 081102 (2018).

\bibitem{Louk} Louk Rademaker and Paula Mellado, ``Charge-transfer insulation in twisted bilayer graphene,'' Phys. Rev. B \textbf{98}, 235158 (2018).
	
\bibitem{Leon1} L.~Balents, ``General continuum model for twisted bilayer graphene and arbitrary  smooth deformations'', SciPost Phys., \textbf{7}, 48 (2019).	

\bibitem{BJYangPRX} J. Ahn, S. Park, and B.-J. Yang, ``Failure of nielsen-ninomiya theorem and fragile topology in two dimensional systems with space-time inversion symmetry: Application to twisted bilayer graphene at magic angle,'' Phys. Rev. X \textbf{9}, 021013 (2019).
	
\bibitem{Bernevig1} Z. Song, Z. Wang, W. Shi, G. Li, C. Fang, and B. Andrei Bernevig, ``All magic angles in twisted bilayer graphene are topological,'' Phys. Rev. Lett. \textbf{123}, 036401 (2019).

\bibitem{Leon2} K. Hejazi, C. Liu, H. Shapourian, X. Chen, and L. Balents, ``Multiple topological transitions in twisted bilayer graphene near the first magic angle,'' Phys. Rev. B \textbf{99}, 035111 (2019).

\bibitem{Dai1} J. Liu, J. Liu, and X. Dai, ``The pseudo-Landau-level representation of twisted bilayer graphene: band topology and the implications on the correlated insulating phase,'' Phys. Rev. B \textbf{99}, 155415 (2019).
	
\bibitem{Fernandes1} J. W. F. Venderbos and  R. M. Fernandes, ``Correlations and electronic order in a two-orbital honeycomb lattice model for twisted bilayer graphene,'' Phys. Rev. B \textbf{98}, 245103 (2018).
	
	
\bibitem{Stauber} J. Gonzalez and T. Stauber, ``Kohn-luttinger superconductivity in twisted bilayer graphene,'' Phys. Rev. Lett. \textbf{122}, 026801 (2019).
	
	
\bibitem{Grisha} G. Tarnopolsky, A. J. Kruchkov, and A. Vishwanath, ``Origin of magic angles in twisted bilayer graphene,'' Phys. Rev. Lett. \textbf{122}, 106405 (2019).

\bibitem{KangVafekPRL} J. Kang and O. Vafek, ``Strong coupling phases of partially filled twisted bilayer graphene narrow bands,'' Phys. Rev. Lett. {\bf 122}, 246401 (2019).
	

\bibitem{Bruno} K. Seo, V. N. Kotov, and B. Uchoa, ``Ferromagnetic Mott state in twisted graphene bilayers at the magic angle,'' Phys. Rev. Lett. \textbf{122}, 246402 (2019).
	
\bibitem{Kaxiras2019} S. Carr, S. Fang, Z. Zhu, and E. Kaxiras, ``Exact continuum model for low-energy electronic states of twisted bilayer graphene'' Phys. Rev. Research, \textbf{1}, 013001, (2019).	

\bibitem{Senthil2} Y.-H. Zhang, D. Mao, Y. Cao, P. Jarillo-Herrero, and T. Senthil, ``Nearly flat chern bands in moire superlattices,'' Phys. Rev. B \textbf{99}, 075127 (2019).
	
\bibitem{Qianghua} Q.-K. Tang, L. Yang, D. Wang, F.-C. Zhang, and Q.-H. Wang, ``Spin-triplet f-wave pairing in twisted bilayer graphene near $\frac14$-filling,'' Phys. Rev. B \textbf{99}, 094521 (2019).	
	
\bibitem{Ashvin1} J. Y. Lee, E. Khalaf, S. Liu, X. Liu, Z. Hao, P. Kim, and A. Vishwanath, ``Theory of correlated insulating behavior and spin-triplet superconductivity in twisted double bilayer graphene,'' Nat. Commun. \textbf{10}, 1 (2019).
	
\bibitem{Roy} B. Roy and V. Juri\ifmmode \check{c}\else \v{c}\fi{}i\ifmmode \acute{c}\else \'{c}\fi{}, ``Unconventional superconductivity in nearly flat bands in twisted bilayer graphene,'' Phys. Rev. B \textbf{99}, 121407 (2019).	

\bibitem{Cantele} P. Lucignano, D. Alf\'{e}, V. Cataudella, D. Ninno, and G. Cantele, ``Crucial role of atomic corrugation on the flat bands and energy gaps of twisted bilayer graphene at the magic angle $\theta\sim 1.08^{\circ}$'', Phys. Rev. B \textbf{99}, 195419 (2019).

\bibitem{Cenke} X.-C. Wu, A. Keselman, C.-M. Jian, K. A. Pawlak, and C. Xu, ``Ferromagnetism and spin-valley liquid states in moire correlated insulators,'' Phys. Rev. B \textbf{100}, 024421 (2019).

\bibitem{Rahaul}  Y.-P. Lin and R. M. Nandkishore, ``Chiral twist on the high-T$_c$ phase diagram in moir\'{e} heterostructures,'' Phys. Rev. B \textbf{100}, 085136 (2019).

\bibitem{SenthilC3} Y. H. Zhang, H. C. Po, and T. Senthil, ``Landau level degeneracy in twisted bilayer graphene: Role of symmetry breaking''
Phys. Rev. B {\bf 100}, 125104 (2019).
	
\bibitem{Ashvin2} S. Liu, E. Khalaf, J. Y. Lee, and A. Vishwanath, ``Nematic topological semimetal and insulator in magic angle bilayer graphene at charge neutrality,'' arXiv:1905.07409.

\bibitem{Sau} Y. Alavirad and J. D. Sau, ``Ferromagnetism and its stability from the one-magnon spectrum in twisted bilayer graphene,'' arXiv:1907.13633.

\bibitem{Dai2} J. Liu and X. Dai, ``Correlated insulating states and the quantum anomalous Hall phenomena at all integer fillings in twisted bilayer graphene'', arXiv:1911.03760.

\bibitem{Fengcheng} F. Wu and S. Das Sarma, ``Collective Excitations of Quantum Anomalous Hall Ferromagnets in Twisted Bilayer Graphene'', Phys. Rev. Lett. \textbf{124}, 046403 (2020).

\bibitem{MacDonald} M. Xie, and A. H. MacDonald, ``Nature of the Correlated Insulator States in Twisted Bilayer Graphene,'' Phys. Rev. Lett. \textbf{124}, 097601 (2020).

\bibitem{Zaletel1} N. Bultinck, S. Chatterjee, and M. P. Zaletel, ``Mechanism for Anomalous Hall Ferromagnetism in Twisted Bilayer Graphene,'' Phys. Rev.
	Lett. \textbf{124}, 166601 (2020).
	
\bibitem{Senthil3Ferro} C. Repellin, Z. Dong, Y.-H. Zhang, T. Senthil, ``Ferromagnetism in narrow bands of moire superlattices,'' Phys. Rev. Lett. \textbf{124}, 187601 (2020).
	
\bibitem{Zaletel2} S. Chatterjee, N. Bultinck, and M. P. Zaletel, ``Symmetry breaking and skyrmionic transport in twisted bilayer graphene,'' Phys. Rev. B \textbf{101}, 165141 (2020).
	
\bibitem{Zaletel3} N. Bultinck, E. Khalaf, S. Liu, S. Chatterjee, A. Vishwanath, and M. P. Zaletel, ``Ground State and Hidden Symmetry of Magic Angle Graphene at Even Integer Filling,'' Phys. Rev. X \textbf{10}, 031034 (2020).
	
\bibitem{Chubukov} D. V. Chichinadze, L. Classen, and A. V. Chubukov, ``Nematic superconductivity in twisted bilayer graphene,'' Phys. Rev. B \textbf{101}, 224513 (2020).
	
\bibitem{YiZhang} Y. Zhang, K. Jiang, Z. Wang, and F. C. Zhang, ``Correlated insulating phases of twisted bilayer graphene at commensurate filling fractions: a Hartree-Fock study,''  Phys. Rev. B \textbf{102}, 035136 (2020).

\bibitem{KangVafekPRB} J. Kang and O. Vafek, ``Non-Abelian Dirac node braiding and near-degeneracy of correlated phases at odd integer filling in magic angle twisted bilayer graphene'', Phys. Rev. B \textbf{102}, 035161 (2020).
	
\bibitem{Guinea2} T. Cea and F. Guinea, ``Band structure and insulating states driven by Coulomb interaction in twisted bilayer graphene'', Phys. Rev. B \textbf{102}, 045107 (2020).	

\bibitem{Ziyang} Y. D. Liao, J. Kang, C. N. Breio, X. Y. Xu, H.-Q. Wu, B. M. Andersen, R. M. Fernandes, and Z. Y. Meng, ``Correlation-induced insulating topological phases at charge neutrality in twisted bilayer graphene'', arXiv:2004.12536.
	
\bibitem{Chubukov2} D. V. Chichinadze, L. Classen, and A. V. Chubukov, ``Orbital antiferromagnetism, nematicity, and density wave orders in twisted bilayer graphene'', arXiv:2007.00871.

\bibitem{Fernandes2} R. M. Fernandes and  J. W. F. Venderbos, ``Nematicity with a twist: rotational symmetry breaking in a moir\'e superlattice,'' Science Advances \textbf{6}, eaba8834 (2020).
	
\bibitem{LeonReview}  L. Balents, C. R. Dean, D. K. Efetov, and A. F. Young, ``Superconductivity and strong correlations in moir\'{e} flat bands'', Nat. Phys. {\bf 16}, 725 (2020).

\bibitem{Lucile} E. Brillaux, D. Carpentier, A. A. Fedorenko, and L. Savary, ``Nematic insulator at charge neutrality in twisted bilayer graphene,'' arXiv:2008.05401.
	
\bibitem{Zaletel4} T. Soejima, D. E. Parker, N. Bultinck, J. Hauschild, and M. P. Zaletel, ``Efficient simulation of moire materials using the density matrix renormalization group'', arXiv:2009.02354.
	
\bibitem{Hunt2017} B. M. Hunt et al., ``Direct measurement of discrete valley and orbital quantum numbers in bilayer graphene'', Nat. Commun.
\textbf{8}, 948 (2017); see Table I in the SI.
		
\bibitem{Niu2020} Y. Ren et al.,  ``WKB estimate of bilayer graphene's magic twist angles'', arXiv:2006.13292.
	
\bibitem{Becker2020} S. Becker et al., ``Mathematics of magic angles in a model of twisted bilayer graphene'' arXiv:2008.08489.
	
\bibitem{NamKoshino2017} N.N.T. Nam and M. Koshino, ``Lattice relaxation and energy band modulation in twisted bilayer graphene'' Phys. Rev. B, \textbf{96}, 075311, (2017).

\bibitem{SM} See Supplemental Material for details of the derivation of the RG equations, RG evolution of the sublattice polarization and Wilson loop eigenvalues, as well as the single particle excitation spectrum in the strong coupling.
	
\bibitem{HerbutPRL2001} I.F. Herbut, ``Quantum Critical Points with the Coulomb Interaction and the Dynamical Exponent: When and Why z=1'' Phys. Rev. Lett. \textbf{87}, 137004 (2001); although there are non-logarithmic corrections to the dielectric {\em function}, whose full consideration is beyond the scope of this work, they do not change the main conclusions.

\bibitem{Watanabe2020} H. Watanabe, ``Counting Rules of Nambu-Goldstone Modes'' Ann. Rev. of Cond. Mat. Phys., \textbf{11}, 169 (2020).

\end{thebibliography}

\begin{thebibliography}{90}

	\bibitem{Leon1} L.~Balents, ``General continuum model for twisted bilayer graphene and arbitrary  smooth deformations'', SciPost Phys., \textbf{7}, 48 (2019).

	\bibitem{Grisha} G. Tarnopolsky, A. J. Kruchkov, and A. Vishwanath, ``Origin of magic angles in twisted bilayer graphene,'' Phys. Rev. Lett. \textbf{122}, 106405 (2019).
	
	\bibitem{KangVafekPRB} J. Kang and O. Vafek, ``Non-Abelian Dirac node braiding and near-degeneracy of correlated phases at odd integer filling in magic angle twisted bilayer graphene'', Phys. Rev. B \textbf{102}, 035161 (2020).

\end{thebibliography}
\end{document}